\documentclass[aps,pra,twocolumn,groupedaddress,amsmath,amssymb,showpacs,showkeys,floatfix]{revtex4-1}

\usepackage{graphicx}% Include figure files
\usepackage{dcolumn}% Align table columns on decimal point
\usepackage{bm}% bold math
\usepackage{amssymb}
\usepackage{multirow}

\usepackage{SIunits}
\usepackage[usenames,dvipsnames]{color}
\newcommand{\mt}[1]{\mathrm{#1}}						% straight letters in Eqs
\renewcommand{\d}{\mt{d}}							% General d for derivatives

\begin{document}

\title{Ground state properties of ultracold trapped bosons with an immersed ionic impurity}

\author{J. M. Schurer}
    \email{jschurer@physnet.uni-hamburg.de}
\author{P. Schmelcher}
\author{A. Negretti}

\affiliation{Zentrum f\"ur Optische Quantentechnologien, Universit\"at Hamburg, Luruper Chaussee 149,
22761 Hamburg, Germany}
\affiliation{The Hamburg Centre for Ultrafast Imaging, Universit\"at Hamburg, Luruper Chaussee 149, 22761
Hamburg, Germany}

\date{\today}


\begin{abstract}

We consider a trapped atomic ensemble of interacting bosons in the presence of a single trapped ion in a 
quasi-one-dimensional geometry. Our study is carried out by means of the newly developed
multilayer-multiconfiguration time-dependent
Hartree method for bosons, a numerical exact approach to simulate quantum many-body dynamics.
In particular, we are interested in the scenario by which the ion is so strongly
trapped that its motion can be effectively neglected. This enables us to focus on the atomic ensemble only.
With the development of a model potential for the atom-ion interaction, we
are able to numerically obtain 
the exact many-body ground state of the atomic ensemble in the presence of an ion. We analyze the influence
of the atom
number and the atom-atom interaction on the ground state properties. Interestingly, for weakly interacting
atoms, we find that the ion impedes the transition from the ideal gas behavior to the Thomas-Fermi limit.
Furthermore, we show that this effect can be exploited to infer the presence of the ion both in the momentum
distribution of the atomic cloud and by observing the interference fringes occurring during an expansion
of the quantum gas. 
In the strong interacting regime, the ion modifies the fragmentation process in dependence of the atom
number parity which allows a clear identification of the latter in expansion experiments.
Hence, we propose in both regimes experimentally viable strategies to assess the impact of the ion on
the many-body state of the atomic gas. This study serves as the first building block for systematically
investigating the many-body physics of such hybrid system.

\end{abstract}

\maketitle
%%%%%%%%%%%%%%%%%%%%%%%%%%%%%%%%%%%%%%%

%%%%%%%%%%%%%%%%%%%%%%%%%%%%%%%%%%%%%%%%%%%%%%%%%%%%%%%%%%%%%%%%%%%%%%%%%%%%%%%%%%%%%%%%
%%% Section I

\section{Introduction}

In recent years, the physics of hybrid atom-ion systems  has attracted more and more attention  both
theoretically and experimentally \cite{Harter2014}. Separately, both systems can be superbly controlled with
 an unprecedented accuracy at the single-particle as well as at the multi-particle level.
However, the combination of atoms and ions poses experimental challenges,
for example, in trapping technology, that have to be overcome. The most prominent one is related to
micromotion, as it hampers the reaching of the ultracold collisional regime for atoms and ions in trapping
systems based on the Paul trap scheme \cite{Cetina2012,Krych2015}. Alternative suggestions to 
circumvent this
problem have
been put forward, for instance, by using optical fields \cite{Schneider2010}. However, it is still an open
question 
which technological solution is the best to accomplish this task. 

Nonetheless, hybrid atom-ion systems make it possible to explore new physics that the two systems 
separately would not
permit. For instance, the development of atom-ion hybrid traps \cite{Smith2005} has paved the way for the
investigation of ultracold elastic and inelastic atom-ion collisions
\cite{Deiglmayr2012,Grier2009,Smith2013,Schmid2010a,Zipkes2011,Haze2013}, controlled chemical reactions
\cite{Hall2012,Ratschbacher2012,Hall2013}, and sympathetic cooling of ions by means of atomic gases
\cite{Willitsch2008,Ravi2012,Hall2012,Smith2005}. Furthermore, such systems lend themselves to applications in
quantum information processing. Exploiting the state-dependent atom-ion interaction allows for the realization
of quantum gates such that the advantages of charged and neutral particles are combined \cite{Doerk2010} or
makes it possible to control
the tunneling in a bosonic Josephson junction such that the generation of entanglement between the atomic
system and
a single ion can be engineered \cite{Gerritsma2012,Joger2014}. Moreover, such systems offer possibilities
to investigate and  understand spin-decoherence processes and spin-exchange interactions at the fundamental
level
\cite{Ratschbacher2013}, aiming at negligible spin relaxation and efficient spin-exchange as desirable
features for quantum information science. 
Besides, atom-ion systems are an excellent platform to simulate
condensed-matter systems and Fr\"{o}hlich polaron Hamiltonians more closely \cite{Casteels2010}. For
instance, an
important component of a solid-state system is the charge-phonon coupling, which is naturally mimicked in an
atom-ion system \cite{Bissbort2013}.
Another interesting application of atom-ion research concerns charge transport in an
ultracold quantum gas. Indeed, it has been shown that a neutral gas doped with few ions should exhibit a
transition from insulating at higher temperatures to conducting at lower temperatures by changing 
the nature of the charge mobility \cite{Cote2000a}.

Apart from these long-term perspectives, up to now  special attention has been paid to hybrid systems realized
in the laboratory  by immersion of a single ion into a Bose-Einstein condensate (BEC)
\cite{Harter2012a,Zipkes2010,Schmid2010a,Zipkes2011}. Theoretical studies of
this setup were considered in the past predicting, for example, the formation of mesoscopic molecular
ions \cite{Cote2002} and ion induced density bubbles  \cite{Goold2010}. However, most of the theoretical
investigations focused on either two or few-body physics or on many-body analyses based on an effective single
particle description, like mean field theory, or by means of a two-mode approximation, like in Ref.
\cite{Gerritsma2012}. To the best of our knowledge, an actual many-body study has been performed only in Ref. 
\cite{Goold2010}. That work, however, focuses on infinite interaction strength, the so-called
Tonks-Girardeau regime \cite{Girardeau1960}, which can be solved exactly
even in the presence of an impurity like an ion. Hence, a detailed
many-body study exploring the weak up to the strong interaction regime in order to
understand the role of the atom-atom interaction, which can be tuned either by Feshbach resonances
\cite{Chin2010} or by modulation of the confinement \cite{Olshanii1998}, is still missing.

Therefore, the aim of the present work is to investigate such a hybrid system in the ultracold regime
from a many-body point of view. To this end, we employ the recently
developed multilayer multiconfiguration time-dependent Hartree method for bosons (ML-MCTDHB)
\cite{Cao2013,Kronke2013}, which is a numerical exact tool to perform time-dependent simulations of many-body
quantum systems. The method belongs to the family of multiconfiguration time-dependent Hartree methods
\cite{Meyer1989,Beck2000} existing for bosons \cite{Alon2008} and fermions \cite{Kato2004}.
We emphasize that the combination of the multilayer structure together with the inclusion of the particle
symmetry of ML-MCTDHB is unique such that it perfectly suits the simulation of the dynamics of such hybrid 
quantum systems. Besides, by means of an improved relaxation method \cite{Beck2000} based on imaginary
time-propagation, ML-MCTDHB makes it possible also to determine stationary states (ground and excited states) 
and their properties.

In particular, we concentrate on the scenario in which
a single strongly trapped ion is immersed in a cloud of ultracold atomic bosons held in a 
quasi-one-dimensional (quasi-1D) trap. Since we consider the
experimentally realistic situation in which the ion is trapped much more tightly than the atoms and since we
assume that the ion is prepared in the ground state of its trap (e.g., by sideband cooling), the ion can be
treated statically. This enables us to neglect the ionic motion. Of course, these assumptions simplify
the simulation of the many-body problem, but we would like to note that this is also the natural first step to
investigate the many-body dynamics in the presence of an impurity ion.

Our analysis focuses mainly on the static properties of this system and the dynamics of the expansion
of the quantum gas. More precisely, we investigate in detail the impact of the ion on the atomic cloud by
comparing the ground state properties to the situation without ion by varying both the atom number and the
atom-atom interaction. Interestingly, for weakly interacting atoms, we find that the ion impedes the
transition from the ideal gas behavior to the so-called Thomas-Fermi (TF) limit. Furthermore, in the strong
interacting regime, the ion modifies the fragmentation process depending on the parity of the number of atoms.
Additionally, we are able to reproduce the density bubbles in a Tonks-Girardeau gas induced by the presence of
the ion as reported in Ref. \cite{Goold2010}. Finally, we show that the presence of the ion manifests itself
in both the momentum distribution of the atomic cloud and the interference fringes occurring during an
expansion. The latter provides us with an indicator to identify the ion and its impact on the atomic cloud in
experiments.

This paper is organised as follows. Section II is devoted to the development of a model potential for the
atom-ion interaction suitable for the subsequent many-body investigations. Indeed, since the ML-MCTDHB method
is based on a spatial grid representation of the many-body wavefunction, it is essential to define the
potential everywhere on the spatial grid. This is not the case for the atom-ion polarization potential
$\sim -r^{-4}$, which is singular at the origin. To this end and for the sake of completeness, we
briefly review the theory of ultracold atom-ion collisions and compare the results obtained via our model
potential to quantum defect theory (QDT). The latter has proven to be an accurate tool for the description of
ultracold atom-ion collisions \cite{Idziaszek2009,Gao2010,Idziaszek2011a,Gao2013}. In Sec. III, we introduce
the many-body Hamiltonian of the interacting atomic ensemble with a centrally localized and static ion in a
quasi-1D setting. In addition, we outline the underlying idea and the most important features of
ML-MCTDHB. In Sec. IV, we present our results of the ground-state properties (i.e., energy, density and
momentum distributions, first- and second-order correlation functions) of the hybrid atom-ion many-body 
quantum system. For the sake of clarity, the section is divided into two main parts. In the first part, we 
investigate weak atom-atom interactions, while in the second one we consider the strong interaction regime. 
In the former case, we analyze the impact of the ion on the transition to the TF regime with a detailed view 
on the atomic density and energy per particle. We draw the connection to the experimental detection of the ion
from the measurement of atomic observables by time-of-flight simulations. For strongly interacting bosons, the
interplay of coherence and fragmentation is investigated up to the fermionization limit in terms of the
reduced density matrices which imprints characteristic features in the time-of-flight behavior. 
Finally, in Sec. V, we conclude our article and give an outlook on future perspectives. 

%%% End Section I
%%%%%%%%%%%%%%%%%%%%%%%%%%%%%%%%%%%%%%%%%%%%%%%%%%%%%%%%%%%%%%%%%%%%%%%%%%%%%%%%%%%%%%%%
%%% Section II

\section{Two-body atom-ion system}

In this section, we briefly describe the ultracold collisions of an atom and an ion with the aim of developing
a suitable model potential for the atom-ion interaction. As we have previously mentioned, this is indeed an
essential ingredient to perform many-body quantum simulations with ML-MCTDHB which we present and
discuss in the subsequent sections. 

\subsection{Atom-ion interaction and 1D conditions} \label{sec:atom-ion}

The interaction between an atom at position $\vec{r}_\mt{A}$ and an ion at position $\vec{r}_\mt{I}$ scales at
large distances as \cite{Seaton1977}
\begin{equation}\label{eq:int3D}
 V_\mt{AI}(r) =  -\frac{C_4}{r^4}.
\end{equation}
Here, $r=|\vec{r}_\mt{A}- \vec{r}_\mt{I}|$ is the inter-particle distance, $C_4 = \alpha e^2/2$, and
$\alpha$ is the static atomic polarizability. It originates from the interaction between the electric field
generated by the ionic charge $e$
and the dipole moment of the atom induced by the ion.  We note that the form given by Eq.
\eqref{eq:int3D} is only valid for
distances larger than a radius $R_0$ which defines the size of the inner core region of the atom-ion
complex. This radius is typically on the order of $2.5 - 4 \angstrom $
\cite{Viehland1994,Cote2000}.
Below $R_0$, the form of the potential is generally unknown.
Additionally, the interaction can be characterized by the  length and energy scale $R^* =
\sqrt{\alpha e^2
\mu/\hbar^2} $ and $E^* = \hbar^2/(2\mu {R^*}^2)$, respectively, where $\mu$ is the reduced mass of the
atom-ion system. Values of $E^*$ and $R^*$ range from a few $\kilo\hertz$ to some hundreds of $\kilo\hertz$
and from tens of $\nano\meter$ to a few hundreds of $\nano\meter$, respectively.
This shows that  the atom-ion interaction is effectively long range, which is opposite to ultracold
atom-atom interactions.

Since we are interested in the simulation of the atom-ion system in a quasi-1D setup, we assume that the
frequency of the transverse confinement is much larger than the
longitudinal one, that is, $\omega_{\bot}\gg \omega_{\parallel}$, for both the atom and the ion.
Transverse trapping frequencies of $\omega_\bot
\approx 2\pi  100\, \kilo\hertz$ can be reached experimentally both for atoms \cite{Moritz2003}
and ions \cite{Schneider2010} leading to a transverse harmonic confinement length of $l_\bot = \sqrt{\hbar /
(\mu\omega_\bot) } \approx 10 - 100 \, \nano\meter$. Without entering into details, it can be shown
\cite{Idziaszek2007} that under these conditions, an effective atom-ion
interaction in the quasi-1D geometry can be derived, whose expression is given by:  
\begin{equation}\label{eq:int1D}
 V_\mt{AI}^\mt{1D}(z) =  -\frac{C_4}{z^4}.
\end{equation}
Here, $z = z_\mt{A} - z_\mt{I}$ denotes the longitudinal atom-ion separation. We note that Eq. 
\eqref{eq:int1D} is valid for $|z| \ge R_\mt{1D}$, where the validity range $R_\mt{1D}$ is defined
by the maximum of the transversal trapping length $l_\bot$ and
the length scale $R_\bot$ at which the polarization potential is equal to the transverse trapping potential.
Since $R^*$ is typically larger than $R_\mt{1D}$, the application of the pseudopotential would be
inappropriate, and therefore we are forced to employ the effective 1D interaction $V_\mt{AI}^\mt{1D}(z)$. This
situation, however, allows us to explore different and more rich physics.
Finally, hereafter we assume always that
the atom and the ion cannot undergo spin-changing collisions, allowing the application of a single-channel
model.

\subsection{Quantum defect theory description} 

Quantum defect theory is a general and powerful method to describe scattering
processes, especially when the pseudopotential approximation is not applicable and the precise form of the
interaction at short distances is unknown.  Its strength stems from its accurate description of the scattering
dynamics by means of  a small number of (energy-independent) parameters, the so-called quantum defect
parameters, which describe the complicated short-range dynamics below interparticle separations of
$R_\mt{1D}$.
We refer the interested reader to Refs. \cite{Greene1982,Seaton1983,Mies1984} for a detailed general
description of QDT and to Refs. \cite{Idziaszek2009,Gao2010,Idziaszek2011a,Gao2013} for its application to
the atom-ion system. 

The ultracold atom-ion scattering in 1D is described by the Schr\"odinger equation 
\begin{equation}\label{eq:SE}
  \left[ -\frac{\hbar^2}{2\mu} \frac{\partial^2}{\partial z^2}  - \frac{C_4}{z^4} \right]\psi(z) = E\psi(z),
\end{equation}
with total energy $E$ and relative wavefunction $\psi(z)$.
For the sake of convenience, we rescale the equation with respect to the characteristic units $R^*$ and
$E^*$, that is, $z \mapsto z/R^*$ and $E \mapsto E/E^*$, such that we obtain
\begin{equation}\label{eq:dimlessSEpaper}
  \left[ \frac{\partial^2}{\partial z^2} + \frac{1}{z^4} + E  \right]\psi(z) = 0.
\end{equation}
The above equation admits both even ($\psi_\mt{e}$) and odd ($\psi_\mt{o}$) solutions and since the atom and
the ion are distinguishable particles, the general solution to Eq. \eqref{eq:dimlessSEpaper}  is given by the
linear combination of $\psi_\mt{e}$ and $\psi_\mt{o}$. 

At short distances, namely when $z\rightarrow
0$, the atom-ion interaction is dominant. Hence, the energy $E$ in Eq. \eqref{eq:dimlessSEpaper} can be
safely neglected and the resulting equation can be solved analytically. Then, the behavior of the relative
wave function at short distances is governed by \cite{Vogt1954}
\begin{align}
 \label{eq:psiEpaper}
 \psi_\mt{e}(z) &\propto |z|\sin(1/|z| + \varphi_\mt{e}), \qquad |z|\ll\sqrt{1/k},   \\
 \label{eq:psiOpaper}
 \psi_\mt{o}(z) &\propto z\sin(1/|z| + \varphi_\mt{o}),  \qquad |z|\ll\sqrt{1/k},
\end{align}
with $k=\sqrt{E}$. These solutions oscillate increasingly fast for small $z$ and in the limit
$|z|\rightarrow 0$ they vanish. The phase of the
oscillation is defined by the quantum defect parameters $\varphi_\mt{e}$ and $\varphi_\mt{o}$ , which depend
on the internal structure of the atom and the ion.
Hence, under the above outlined assumptions for the validity of Eq. \eqref{eq:dimlessSEpaper}, the
quantum defect parameters effectively set boundary conditions at $|z| = R_\mt{1D}$ (in numerical practice
when $z\rightarrow 0$).  Besides, we note that in the low
energy limit, the quantum defect
parameters $\varphi_\mt{e}$ and $\varphi_\mt{o}$ are related to the 1D zero energy scattering lengths as
$a_\mt{e,o}(k=0) =-\cot{(\varphi_\mt{e,o})}R^*$.
Since experimental values for the scattering lengths are still not available, we use $\varphi_\mt{e,o}$ as
adjustable parameters in the range $[-\pi/2 , \pi/2]$.

In the opposite  limit $|z|\gg R^*$, the solutions of Eq. \eqref{eq:dimlessSEpaper} for $E>0$
are given by plane waves 
\begin{equation} \label{eq:psiAsym}
 \psi^\mt{asym}_\mt{e,o} \propto \sin{(kz)} + \tan{(\xi_\mt{e,o})}\cos{(kz)}
\end{equation}
with asymptotic phase shifts $\xi_\mt{e,o}$.
As shown in Ref. \cite{Idziaszek2011a}, these asymptotic phase shifts and the scattering lengths, 
$a_\mt{e,o}(k)=-\tan(\xi_\mt{e,o})/k$, can be analytically related to the
quantum defect parameters.
This connection can be viewed as the full solution of the scattering problem. Please note that this QDT
formalism is indeed applicable in 1D when the condition $l_{\perp} \gg R_0$ applies, which is generally
fulfilled.

\begin{figure}
 \includegraphics[width = 1.0\linewidth]{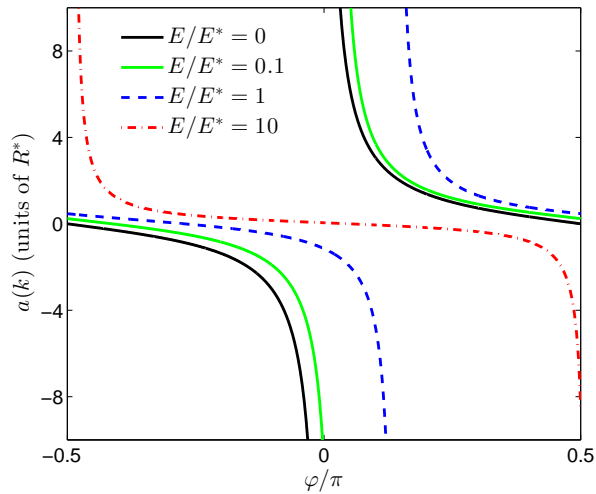}
 % phi_20.eps: 0x0 pixel, 300dpi, 0.00x0.00 cm, bb=   18   179   594   612
 \caption{ (Color online) Scattering length $a(k)/R^*$ versus the quantum defect parameter $\varphi$ for four
different energies $E/E^* = 0 , 0.1 , 1 , 10 $. Since this relation does not depend on the parity of the
solution $\psi(z)$, we dropped the subscript $\mt{e,o}$.}
 \label{fig:phaseShift}
\end{figure}

In Fig. \ref{fig:phaseShift}, we illustrate the dependence of  $a_\mt{e,o}(k)$ on $\varphi_\mt{e,o}$ for
different energies $E$. This relationship will turn out to be useful in Sec. \ref{ssec:Connect} when
connecting our atom-ion model potential to QDT. We can observe that the scattering length can be
tuned from $-\infty$ to $\infty$ by adjusting the quantum defect parameters and that it strongly depends on
the total energy $E$. This shows that the modeling with the pseudopotential is indeed inappropriate.

\begin{figure}
 \includegraphics[width = 1.0\linewidth]{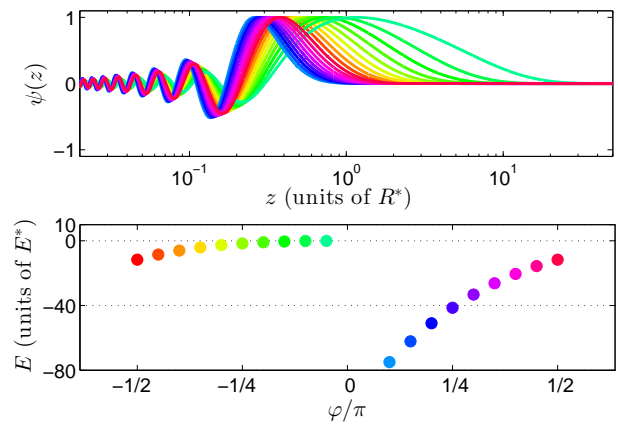}
 % phi_20.eps: 0x0 pixel, 300dpi, 0.00x0.00 cm, bb=   18   179   594   612
 \caption{(Color online) Most weakly bound states of the polarization potential. In the
top panel, the wavefunctions for different quantum defect parameter are shown. Since even and odd solutions
are equal for $z>0$, we show only the positive semi-axis and dropped the index $\mt{e,o}$. The full
solutions can be found by either symmetric or antisymmetric continuation to $z<0$.
In the lower panel, the corresponding  energies are shown as a function of the quantum defect parameters
$\varphi_\mt{e,o}$. To identify the energy of a    
certain wavefunction, they are colored corresponding to their energy.}
 \label{fig:boundStates}
\end{figure}

Finally, we conclude with a brief discussion concerning the bound states supported by the polarization 
potential. To begin with, we note that the number of bound states depending on the cutoff 
radius $R_\mt{1D}$ can be estimated via the following relation \cite{Massignan2005}:
\begin{equation}\label{eq:nofBoundStates}
 \nu = \mt{Int}\left[ R^*/(R_\mt{1D} \pi) \right].
\end{equation}
For example, for $R_\mt{1D}=0.02 R^*$ we have 15 bound states. Since typically most of these states are
strongly
localized and far detuned from the energy threshold, only the weakest bound states are of relevance for
the ultracold atom-ion collision. In Fig. \ref{fig:boundStates}, we show the weakest bound states together
with their energy as a function of the quantum defect parameters $\varphi_\mt{e,o}$. In order to compute them
we used the renormalized Numerov method \cite{Czuchaj1987,Johnson1977}. Two main observations can be made:
The spatial maximum of the bound
state varies in position from $|z| = 0.2 R^*$ up to $|z| = R^*$; its energy changes from near
threshold for $\varphi_\mt{e,o} < 0$ to strong binding energies of about $E=-80 E^*$ for $\varphi_\mt{e,o} >
0$. 

Given this, we now have the necessary background information to construct a proper model
potential for the atom-ion interaction, which is the topic of the next section. 

\subsection{Model potential} 

In a many-body theory, we can not use, as already mentioned, an interaction potential in the form of Eq.
\eqref{eq:int1D} due to the need for boundary conditions originating from the limited validity range.
Therefore, we introduce a model potential defined for all values of $z$.

Such a model potential has to fulfill three criteria. First, it should reproduce the
$-1/z^4$ long-range tail in order to result in correct bound (at least some) and scattering solutions. Second,
it needs to regularize the unphysical divergence occurring due to the limited validity range  of the
polarization potential. The regularization will also help to reduce the increasingly fast oscillations for
$z\rightarrow 0$, which would be numerically very difficult to handle in ML-MCTDHB. Third, the boundary
conditions imposed by the
QDT [Eqs. \eqref{eq:psiEpaper} and \eqref{eq:psiOpaper}] have to be included in a flexible way such that most
of the quantum defect parameter combinations $\{\varphi_\mt{e},\varphi_\mt{o} \}$ can be indeed modeled. 

A good choice fulfilling the above outlined criteria is given by the following model potential:
\begin{equation}
 \label{eq:modelPotpaper}
 V_{\mt{mod}}(z) = \mt{v}_0 e^{-\gamma z^2} - \frac{1}{z^4 + 1/\omega}.
\end{equation}
The polarization tail is controlled by the parameter $\omega$, expressed in units of $(R^*)^{-4}$, which
can be understood as the potential depth.
Indeed, for $|z| \rightarrow 0$,  the second term in Eq.
\eqref{eq:modelPotpaper} approaches $-\omega$. Additionally, $\omega$ defines the number of bound states
within the model potential. Moreover, we superimpose a Gaussian barrier at $z=0$ which is characterized by the
height $\mt{v}_0 > 0$, in units
of $E^*$, and the inverse width $\gamma$ in units of $(R^*)^{-2}$.
Its purpose is to model the short-range behavior of the polarization potential by properly varying $\mt{v}_0$
and $\gamma$ such that several quantum defect parameters $\varphi_\mt{e,o}$ can be modeled.
Note that the Gaussian should be localized in the spatial region dominated by the parameter $\omega$ ($z^4 \ll
1/\omega$)  in order to prevent it from disturbing the long range part of the potential.
This is achieved by setting a
minimal value for $\gamma$ in dependence of $\omega$ in the following way. Equating the Gaussian $2\sigma$
range ($2\sqrt{1/\gamma}$) to the length scale at which $\omega$ dominates the $-1/z^4$
polarization tail [$\approx\sqrt[4]{1/(10\omega)}$]  leads to the restriction $\gamma \geq 4
\sqrt{10\omega}$.
Besides, we note that the Gaussian height $\mt{v}_0$ has to be large enough in order to generate a
vanishing wavefunction at $z=0$, and to this end we set $\mt{v}_0 = 3\omega$.

\begin{figure}
 \includegraphics[width = 1.0\linewidth]{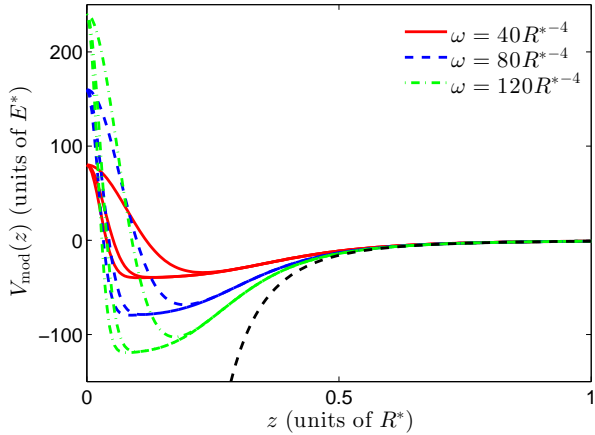}
 % phi_20.eps: 0x0 pixel, 300dpi, 0.00x0.00 cm, bb=   18   179   594   612
 \caption{(Color online) Model potential for the parameter combinations $\omega= [40, 80, 120] /(R^*)^{4}$
and $ \gamma = [1 , 5 , 9 ]\gamma_\mt{min}$ with $\gamma_\mt{min} = 4\sqrt{10\omega}$.
Potentials with the same $\omega$ are plotted in equal color (line style). For each $\omega$ three different
values of $\gamma$ are shown. Small values of $\gamma$ correspond to a broad Gaussian, while large
ones correspond to a narrow Gaussian. For clarity, we also plot the polarization potential $-1/z^4$ (black
dashed line). }
 \label{fig:modPotpaper}
\end{figure}

In Fig. \ref{fig:modPotpaper}, we show an example of the model potential for several parameter combinations
$\{\omega , \gamma \}$. As it is displayed, good agreement between $V_\mt{mod}(z)$ and the $-1/z^4$
behavior for separations beyond $|z|>0.5 R^*$ is achieved. Furthermore, we can see that for larger $\omega$
the model potential approaches more and more the $-1/z^4$ curve, even for smaller inter-particle distances,
such that we can view $\omega$ as the parameter controlling the degree of approximation.

\subsection{Connecting the model potential and quantum defect theory}\label{ssec:Connect}

Let us next connect the parameters characterizing $V_\mt{mod}(z)$ to the quantum defect parameters in order to
demonstrate to what extend our model potential is capable to describe the ultracold atom-ion scattering
process.
This will enable us to establish a mapping $\{\varphi_\mt{e},\varphi_\mt{o}\} \leftrightarrow \{\omega,\gamma
\}$ by comparing the asymptotic phase shifts $\xi_\mt{e,o}$ [see Eq. \eqref{eq:psiAsym}]
of QDT to those of the model potential.

In order to derive the model solutions $\psi^\mt{mod}(z)$,  the
interaction term $-1/z^4$ in Eq. \eqref{eq:dimlessSEpaper} is replaced by the model potential, and we
solve the resulting equation again by means of the Numerov method. As boundary conditions, we use
for the even solutions $\mt{d}\psi^\mt{mod}_\mt{e}(z)/\mt{d}z \xrightarrow {z\rightarrow 0} 0$ and for the odd
solutions $\psi^\mt{mod}_\mt{o}(z) \xrightarrow{z\rightarrow 0} 0$.
 For these model solutions with energy $E=k^2$, the phase shift can be obtained by
comparing the logarithmic derivative of $\psi^\mt{mod}(z)$ and a plane-wave solution at position $d\gg R^*$.
 This yields the relation 
\begin{equation}\label{eq:phaseShift}
 \cot{\left( \xi_\mt{e,o}\right) } = \frac{k+A_\mt{e,o}\cot{\left( kd\right) }}{-A_\mt{e,o} + k\cot{\left(
kd\right) }}
\end{equation}
with $ A_\mt{e,o} = \frac{\d\psi^\mt{mod}_\mt{e,o}(z)/\d z}{\psi^\mt{mod}_\mt{e,o}(z)} \big|_{z=d}$. Since the
relation between the quantum
defect parameters and  the phase shift $\xi_\mt{e,o}$ is known analytically for the pure polarization
potential from QDT (see also Fig. \ref{fig:phaseShift}), we can match the asymptotic phase shifts, and
therefore obtain the desired mapping.

\begin{figure}
 \includegraphics[width = 1.0\linewidth]{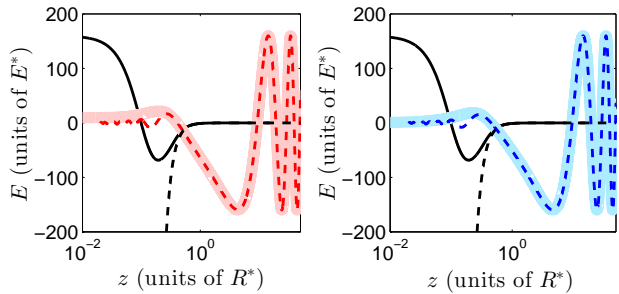}
 % phi_20.eps: 0x0 pixel, 300dpi, 0.00x0.00 cm, bb=   18   179   594   612
 \caption{(Color online) Scattering solutions for $E=0.1 E^*$. Even (left panel) and odd (right panel)
solutions of the model potential for $\omega=80 (R^*)^{-4}$ and $\gamma=\gamma_\mt{min}$ are shown with
thick solid lines. The corresponding QDT solutions with quantum defect parameters $\varphi_\mt{e} =
0.23\pi$ and $\varphi_\mt{o}=0.3\pi$, obtained by the mapping (see text), are show as dashed thin lines
in the corresponding color. The solid black line
represents the corresponding model potential and the black dashed line the polarization potential. Note that
due to the symmetry of the solutions it is sufficient to show the positive semi-axis.}
 \label{fig:scatteringCompare}
\end{figure}

In Fig. \ref{fig:scatteringCompare}, we show examples of the even (left panel) and the odd (right panel)
solutions (thick solid lines) for $E = 0.1 E^*$, $\omega = 80 (R^*)^{-4}$, and $\gamma= \gamma_\mt{min}$ 
and, to best visualize the short-range behavior, we use a logarithmic scale for the $z$ axis. Additionally,
the corresponding QDT solutions (dashed thin lines) are shown. We can observe that both solutions coincide
perfectly for $|z|>0.2 R^*$. Only in the vicinity of $z=0$ the model solutions approximate the QDT results in
the manner of an envelope. This shows that our model is indeed capable to reproduce the QDT scattering
behavior quite accurately.

\begin{figure}
 \includegraphics[width = 1.0\linewidth]{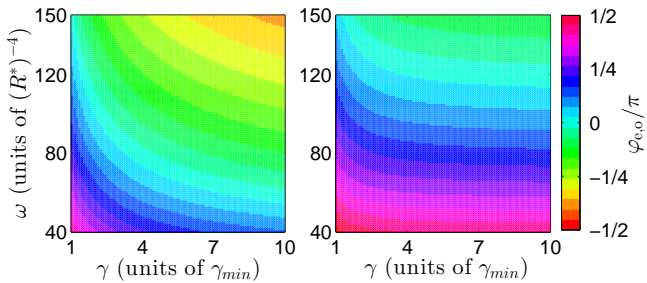}
 % phi_20.eps: 0x0 pixel, 300dpi, 0.00x0.00 cm, bb=   18   179   594   612
 \caption{(Color online) Mapping $ \{\omega , \gamma \} \rightarrow
\{\varphi_\mt{e},\varphi_\mt{o} \}$ between our model potential and 1D QDT: Even (left panel) and odd
(right panel) solutions. }
 \label{fig:mapping}
\end{figure}

We have carried out such an analysis in a systematic way in order to identify the mapping for all values of
the parameters of the model potential.
The result is shown in Fig. \ref{fig:mapping} for the even (left panel) and the odd (right panel) solutions
where the color encodes the values of the quantum defect parameters $\varphi_\mt{e,o}$. The
range of $\omega$ is chosen in such a way that the model potential has two bound states and $\gamma$ varies
from its minimal value $\gamma_\mt{min} = 4\sqrt{10\omega}$ to $10 \gamma_\mt{min}$. 

We observe that a wide range of the quantum defect parameters $\varphi_\mt{e,o}$ can be modeled. Moreover,
the odd quantum defect parameter $\varphi_\mt{o}$ is nearly independent of $\gamma$ such that it can be tuned
by the parameter $\omega$. The parameter $\gamma$ can then be used to set the even quantum defect parameter
$\varphi_\mt{e}$ independently of $\varphi_\mt{o}$.
This fact is helpful for finding the desired model potential more easily. 
Additionally, we would like to point out that the mapping $  \{\omega , \gamma \}  \rightarrow
\{\varphi_\mt{e},\varphi_\mt{o} \}$ shown in Fig. \ref{fig:mapping} is independent of the energy $E$. This
is a necessary requirement in order to use it as a connection between the two parameter spaces.

In conclusion, we have established a mapping which makes it possible to determine the model parameters for a
given pair of quantum defect parameters. Besides, we have shown that our choice of model potential can
reproduce very well the results of QDT. Given this, we can now perform efficiently
many-body quantum simulations, in particular, in a regime beyond the pseudopotential approximation. 

%%% End Section II
%%%%%%%%%%%%%%%%%%%%%%%%%%%%%%%%%%%%%%%%%%%%%%%%%%%%%%%%%%%%%%%%%%%%%%%%%%%%%%%%%%%%%%%%
%%% Section III

\section{Hamiltonian and computational approach}

In this section, we first present the many-body system we are going to investigate, namely, a single
strongly localized ion immersed in an interacting atomic cloud in 1D. Afterwards, we briefly review the
ML-MCTDHB which we use to
compute the many-body wavefunction and its time evolution. We would like to stress once again that in our
current study the ion is treated statically, and therefore no ionic motion is considered. We show
that already for a fixed ion intriguing phenomena and structures do appear for the atom-ion compound.

\subsection{Many-body atom-ion system}

Hereafter, we consider a hybrid system consisting of a single static ion and a cloud of $N$ bosonic
atoms at zero temperature in the quasi-1D regime. As already discussed in Sec. \ref{sec:atom-ion}, for 
both systems this
situation can be met experimentally.
Besides, we also assume that the static ion is localized in the center of the trap ($z=0$) confining the
atoms, which we consider to be harmonic.
We note that in order to treat the ion statically, the ionic trapping frequency has to be larger than
the frequency of the atomic trap such that the
ionic wave function is strongly localized. However, in an experimental realization based on a radio-frequency
trap for the ion the effect of micromotion can be quite important for atom-ion systems, as
extensively discussed in Ref. \cite{Nguyen2012}. Nevertheless,
there it has been found that bound atom-ion states, are almost unaffected by the
micromotion. Hence, since we focus in the following mainly on the properties of the ground state of the
system, which is principally dominated by bound states, the effect of micromotion can be safely neglected
in our subsequent analysis.

Now the interaction between the atom at position $z_j$ and the ion at position $z_\mt{I} = 0$ is described by
our previously introduced model potential $V_\mt{mod}(z_j)$ [see Eq. \eqref{eq:modelPotpaper}]. Further, we
allow for interactions among the atoms. Due to the short-range nature of the ultracold atom-atom scattering,
we model their interaction by a pseudopotential. Hence, the quasi-1D
bosonic gas with a central static ion is described by the Hamiltonian
\begin{equation}\label{eq:Hamiltonian}
  \hat{H} = \sum_i^N \left[ - \frac{\partial^2}{\partial z_i^2} + V_\mt{trap}(z_i) + V_{\mt{mod}}(z_i) 
\right] 
+                  g\sum_{i<j}^N  \delta(z_i-z_j)
\end{equation}
which is expressed in new characteristic 
polarization length and energy units $R^* = \sqrt{\alpha e^2 m/\hbar^2} $ and $ E^* = \hbar^2/(2m {R^*}^2)$,
respectively. Here, we have replaced the reduced mass $\mu$ used
in the relative frame with the mass $m$ of the atoms \footnote{We make this choice only for numerical
convenience, as the use of the old units introduced in Sec. II would have the effect to introduce an
overall factor $\mu/m$. Of course, this transformation does not change the underlying physics.}.
The first sum appearing in Eq. \eqref{eq:Hamiltonian} contains the single
particle operators: the kinetic energy, the longitudinal harmonic trap 
\begin{equation}\label{eq:harmonic}
 V_\mt{trap}(z) = \frac{1}{l_\parallel^4}z^2
\end{equation}
with trapping length $l_\parallel = \sqrt{\hbar/(m\omega_\parallel)}/R^*$ expressed in units of $R^*$, and the
ionic potential. The second sum
represents the atom-atom interaction with the coupling constant $g$ expressed in units of $E^* R^*$.

In order to easily resolve the polarization potential within our numerical calculations, we choose
$R^*$ to be of the order of $l_\parallel$, e.g.,  $l_\parallel = 1/2$. Assuming $\omega_\parallel \approx 
2\pi 
 1\,\kilo\hertz$, this choice, for instance, corresponds to
$R^* \approx 400\,\nano\meter$ and $l_\parallel \approx 200\, \nano\meter$  for $^{87}$Rb atoms.
Further, we fix for the remaining part of the paper the quantum defect parameters to $\varphi_\mt{e} =
0.23\pi$ and $\varphi_\mt{o} = 0.3 \pi$ which correspond to the model parameters $ \omega = 80 (R^*)^{-4}$
and $\gamma = \gamma_\mt{min}$. We note, however, that this choice for the quantum defect parameters is not
essential for the ground-state properties of the quantum gas we present later in the paper. Nevertheless,
the choice of $\varphi_\mt{e,o}$ can be relevant for the dynamics, as it can for example lead to assistance or
inhibition of tunneling in a bosonic Josephson junction \cite{Gerritsma2012}.

\subsection{Methodology}\label{secc:method}

The ML-MCTDHB belongs to the class of multiconfiguration time-dependent
Hartree methods. They all share the concept to expand the many-body wavefunction in a time-dependent
basis which is comoving with the system. This allows for an accurate and numerically efficient
simulation of the dynamics of an interacting quantum many-body system. In the extension for bosons, the
many-body wavefunction is symmetrized in order to respect the bosonic symmetry of the particles. The
multilayer feature enables us to even simulate bosonic mixtures taking all (inter- and intraspecies)
correlations into account \cite{Cao2013,Kronke2013}.

The ansatz for the many-body wave function is taken as a linear combination of Hartree products built by
$M_\sigma$ states $|\psi_i^{(\sigma)}(t)\rangle$ ($i=1\cdots M_\sigma$) for each species $\sigma$:
\begin{equation}
 |\psi(t)\rangle = \sum_{i_1=1}^{M_1} \cdots \sum_{i_S=1}^{M_S} A^1_{i_1 \cdots i_S}(t)
|\psi_{i_1}^{(1)}(t)\rangle \cdots |\psi_{i_S}^{(S)}(t)\rangle.
\end{equation}
The coefficients on this first layer of the expansion, denoted by $A^1_{i_1 \cdots i_S}(t)$, depend on the
species indices  $i_\sigma$ and on time $t$. In a hybrid atom-ion setup, we would have two species ($S=2$)
with the first $\sigma=1$ representing the atoms and the second $\sigma=2$ the ion.

On the second layer, the species wavefunctions for $N_\sigma$ bosons are expanded in bosonic number states
$|\vec{n}\rangle_t^\sigma$ to incorporate their indistinguishability,
\begin{equation}
 |\psi_i^{(\sigma)}(t)\rangle = \sum_{\vec{n}|N_\sigma} A_{i;\vec{n}}^{2;\sigma}(t) |\vec{n}\rangle_t^\sigma.
\end{equation}
In a number state $|\vec{n}\rangle_t^\sigma$, each $\sigma$ boson occupies one of the $m_\sigma$ single
particle functions (SPFs)
$|\phi_j^{(\sigma)}(t)\rangle$. The vector $\vec{n} = (n_1,\cdots,n_{m_\sigma})$ contains the occupation
numbers $n_j$ of every SPF. Further, the coefficients $A_{i;\vec{n}}^{2;\sigma}(t)$ depend on the
species $\sigma$, on the species number state $\vec{n}$, and on time $t$.

On the third and last layer, the time-dependent SPFs are
represented in a time-independent basis $\{|r_j^\sigma\rangle\}_{j=1}^{\mathcal{M}_\sigma}$,
\begin{equation}
 |\phi_j^{(\sigma)}(t)\rangle = \sum_{i=1}^{\mathcal{M}_\sigma} A^{3;\sigma}_{j;i}(t) |r_i^\sigma\rangle
\end{equation}
with the coefficients $A^{3;\sigma}_{j;i}(t)$.

This expansion of the full many-body wave function is based on a cascade of truncations introduced
through finite basis sets $\{|\psi_j^{(\sigma)}(t)\rangle\}_{j=1}^{M_\sigma}$,
$\{|\phi_j^{(\sigma)}(t)\rangle\}_{j=1}^{m_\sigma}$, and  $\{|r_j^\sigma\rangle\}_{j=1}^{\mathcal{M}_\sigma}$,
leading to a wave function in a truncated Hilbert space. This truncation error can be kept small even during
the dynamical evolution of the system due to the time-dependent basis functions.

By using the Dirac-Frenkel variational principle \cite{Dirac1930,Frenkel1934},
\begin{equation}
 \langle \delta\psi|  ( i\partial_t -\hat{H} )|\psi\rangle = 0,
\end{equation}
where  $\langle \delta\psi|$ denotes the variation of the total wave function, one can derive the equation of
 motion for the above outlined expansion coefficients on each layer in order to 
describe the temporal evolution of the wave function. Please note that the usage of the Dirac-Frenkel
variational principle guarantees variational optimal basis sets which makes it possible to keep the number of 
needed basis functions small. We refer to Refs. \cite{Cao2013,Kronke2013} for
more details.

We note that in addition to the real-time evolution, ML-MCTDHB makes it possible to obtain the ground and the 
excited states of the system by  imaginary time propagation \cite{Kosloff1986} of an initial guess wave 
function.
In the simplest case, such an initial wave function can be a number state built by the non interacting 
single-particle functions $|\phi_i^0\rangle$ with energies $E_i^0$.

The analysis of the resulting high-dimensional time-dependent many-body
wave function is typically carried out in terms of reduced density matrices.
The one-particle density matrix and its spectral decomposition can be written as
\begin{equation}
 \label{eq:natpopOrb}
\rho_1 = N tr_{2\cdots N} |\psi\rangle \langle \psi|= \sum_i n_i |\chi_i \rangle \langle\chi_i|.
\end{equation}
Here $|\chi_i\rangle$ are the so-called natural orbitals and $n_i$ represent the natural populations with the
property $\sum_i n_i = N$. With these $n_i$, we can judge the convergence of the algorithm
\cite{Beck2000}. Furthermore, the largest natural population $n_0$ can be used as a measure of the
fragmentation \cite{Penrose1956}. One refers to a fragmented state or fragmented condensate, when more than
one natural population is on the order of $N$. Fragmentation in the ground state is especially linked to the
interacting strength, since to minimize the interaction energy it is favorable to distribute the
particles in many natural orbitals \cite{Pethick}.
When $n_0/N=1$, one recovers the well-known Gross-Pitaevskii solution and therefore $|\chi_0 \rangle$ is also
sometimes referred to as the BEC state.

Finally, since we will also investigate two-body correlations, we simply recall that these are described by
the diagonal elements of the reduced two-body density matrix $\rho_2$, which is defined as
\begin{equation}
 \label{eq:g2}
  \rho_2(r,r') =   \langle r,r^\prime|  \left( tr_{3\cdots N} |\psi\rangle \langle
\psi| \right)  |r,r^\prime\rangle .
\end{equation}
It represents the probability of finding a boson at position $r$ and another one at position $r^\prime$. 

%%% End Section III
%%%%%%%%%%%%%%%%%%%%%%%%%%%%%%%%%%%%%%%%%%%%%%%%%%%%%%%%%%%%%%%%%%%%%%%%%%%%%%%%%%%%%%%%
%%% Section IV

\section{Results}

In this section, we analyze in detail the ground-state properties of the system described above.
We separate the investigation into two
parts: weakly and strongly interacting bosons up to the fermionization regime. 

For the subsequent analyses, the so-called Lieb-Liniger parameter $\gamma_{LL}$ \cite{Lieb1963} turns out to
be a useful measure for the degree of the interactions. We recall that for a homogeneous system in 1D the
effective interaction strength can be described  by $\gamma_{LL} = g/(2\rho)$, which is the ratio
between the
interaction strength $g$ and the mean density $\rho = N/L$, where $L$ is the system size. For $\gamma_{LL} \ll
1$, we have the weakly interacting or equivalently high-density regime, which is characterized by single
particle behavior. In
contrast, $\gamma_{LL} \gg 1 $ corresponds to the strongly interacting or low-density regime revealing
fermionization behavior for $ \gamma_{LL} \rightarrow \infty$ \cite{Girardeau1960}.

In order to analyze the impact of the ionic potential on the bosonic cloud, we perform all calculations with
and without the ion potential such that we can compare the case of the purely harmonic trapping (i.e., without
ion) to the situation of the additional ionic potential. In the following, we refer to these two situations as
the harmonic (H) case and the ionic (I) case, respectively, to distinguish between the two different 
scenarios.

\subsection{Weak interactions}

To begin with, we analyze the spatial
density distribution $n(z)$ and the energy per particle $E/N$. Thereby, it is convenient to separate  the
energy into its components\textemdash kinetic, trapping, interaction, and ionic\textemdash such that  $E = 
E_{\mt{kin}}+E_{\mt{trap}} + E_{\mt{int}} + E_\mt{ion}$.
Besides, we fix the interaction strength to the value $g=2 E^*R^*$ and vary the number of bosons from $N=2$ up
to $N=200$. 
Such a choice for $g$ well describes the regime in which we are now interested because the population of the
lowest natural orbital $n_0/N$ does not fall below $95\%$ for all $N$. This becomes even more the case for
large particle numbers because the Lieb-Liniger parameter scales as $1/N$ (for $N=50$ we have $n_0/N = 0.98$).
Therefore, the system could be described to a good approximation by the Gross-Pitaevskii equation,
where the atoms occupy only one single-particle orbital $|\chi_0\rangle$.
Given this, we use only $m=3$ SPFs in our numerical simulations in this regime.

\subsubsection{Energy in the low- and high-particle-number regimes}

One can distinguish the low- and the high-particle-number regimes.
For small numbers of particles, the system should behave like an ideal gas because the interaction
energy $E_\mt{int}$ is small with respect to $E$. Thus, we can expect the
density to be given by the lowest non-interacting single-particle orbital
occupied by all particles, that is,  $n_\mt{ideal}(z) = N|\phi^0_0(z)|^2$. Then the ground-state energy of
the system can be estimated by using first-order perturbation theory, with respect to the interaction term
appearing in Eq. \eqref{eq:Hamiltonian}, yielding
\begin{equation}\label{eq:GSGPlowNenergy}
  \frac{E_\mt{weak}(N)}{N} = E_0^0 + \frac{g}{2}(N-1)\int{ |\phi^0_0(z)|^4 \d z}.
\end{equation}
In the harmonic case, this leads to a ground-state energy per particle of
$E_\mt{weak}^\mt{H}(N)/N = E_0^0 + g(N-1)/(\sqrt{2\pi}l_\parallel)$ scaling linearly with the particle number.
In the presence of the ion, Eq. \eqref{eq:GSGPlowNenergy} has to be evaluated numerically, but the linear
scaling with $N$ is still preserved.

\begin{figure}
 \includegraphics[width = 0.6\linewidth]{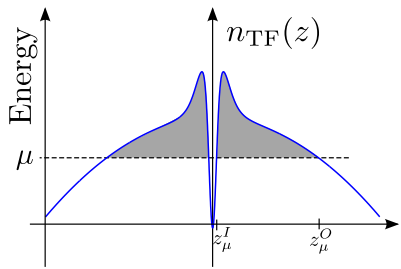}
\caption{(Color online) Sketch of the TF density $n_\mt{TF}(z)$ with the ion. The blue line
corresponds to $-V(z)$, where $V(z)$ is the sum of the trapping and the atom-ion potentials. The black dashed
line marks the chemical potential $\mu$. Above $\mu$, the
condition $n_\mt{TF}(z)>0$ holds and when $\mu=V(z)$ an inner  $z_\mu^I$ and outer $z_\mu^O$ TF radius
can be defined leading to the density distribution $n_\mt{TF}(z)$ shown by the gray shaded area.}
\label{fig:TFcartoon}
\end{figure}

On the other hand, for large particle numbers, the system is well-described by the TF
approximation. In this limit, the density can be described by
$n_\mt{TF}(z) = (\mu-V(z))/g$, where $V(z)$ is either only the trap or the sum of the trap and the ionic
potential, whereas  $\mu$ denotes the chemical potential.
Thus, the total energy can be expressed in the TF regime as
\begin{equation}\label{eq:GSTFenergy}
 E_\mt{TF}(N) = \int\limits_{\{n_\mt{TF}(z)>0\}} \d{z} \left[ V(z)n_\mt{TF}(z) + \frac{g}{2} (n_\mt{TF}(z))^2
\right] .
\end{equation}
For the harmonic case, following Ref. \cite{Pethick}, one can analytically derive the energy per
particle: $ E_\mt{TF}^\mt{H}(N)/N  = 3 e_0 N^{2/3}/5 $ with
$e_0=[3g/ (4l_\parallel^2) ]^{2/3}$, and the TF radius $z_\mu^\mt{H} = l_\parallel^2 \sqrt{\mu}$. In order to
understand the composition of the total energy, the ratio of
interaction energy and potential energy can be derived leading to $E_\mt{int}^\mt{H}/E_\mt{trap}^\mt{H} = 2$.
For the ionic case, we obtain $E_\mt{TF}(N)$ numerically by imposing the condition $\mu=g n(z_0)$ with 
$V(z_0)= 0$ from which one can define an inner $z_\mu^I$ and an outer $z_\mu^O$ TF radius. This defines a
zone in which $n_\mt{TF}(z)$ is well-defined as it is indicated by the gray region in
Fig. \ref{fig:TFcartoon}. 

Now, for large particle numbers and for the ionic case, Eq. \eqref{eq:GSTFenergy} can be approximated, 
yielding for the total energy the following expression:
\begin{equation}\label{eq:TFion}
 E_\mt{TF}^\mt{I}(N)  = \frac{3 e_0}{5} (N-N_0)^{5/3} + E^\mt{I}_0.
\end{equation}
Here $N_0 = -g^{-1} \int_{-z_\mu^O}^{z_\mu^O} V_\mt{mod}(z) \d{z}  $ and $ E^\mt{I}_0 = -g^{-1}
\int_{-z_\mu^O}^{z_\mu^O} [ V_\mt{trap}(z)V_\mt{mod}(z) +  V_\mt{mod}(z)^2 /2] \d{z}  $. Expressions for the
components of the energy per particle are given in the Appendix \ref{app:TFIon}.
Similarly, one can show that for $N\rightarrow \infty$
the ratio between the interaction energy and the potential energy is the same as in the harmonic case, i.e.,
$E_\mt{int}^\mt{I}/(E_\mt{trap}^\mt{I}+E_\mt{ion}^\mt{I}) = 2$.  Therefore, we can conclude that for 
$N\rightarrow \infty$ (i.e.,
 $N\gg N_0$) the impact of the ion on the interacting atomic cloud becomes negligible even though the
shift by $N_0$ of the total energy does not vanish.

\subsubsection{Density and energy per particle in the harmonic and the ionic cases}

\begin{figure}
 \includegraphics[width = 1.0\linewidth]{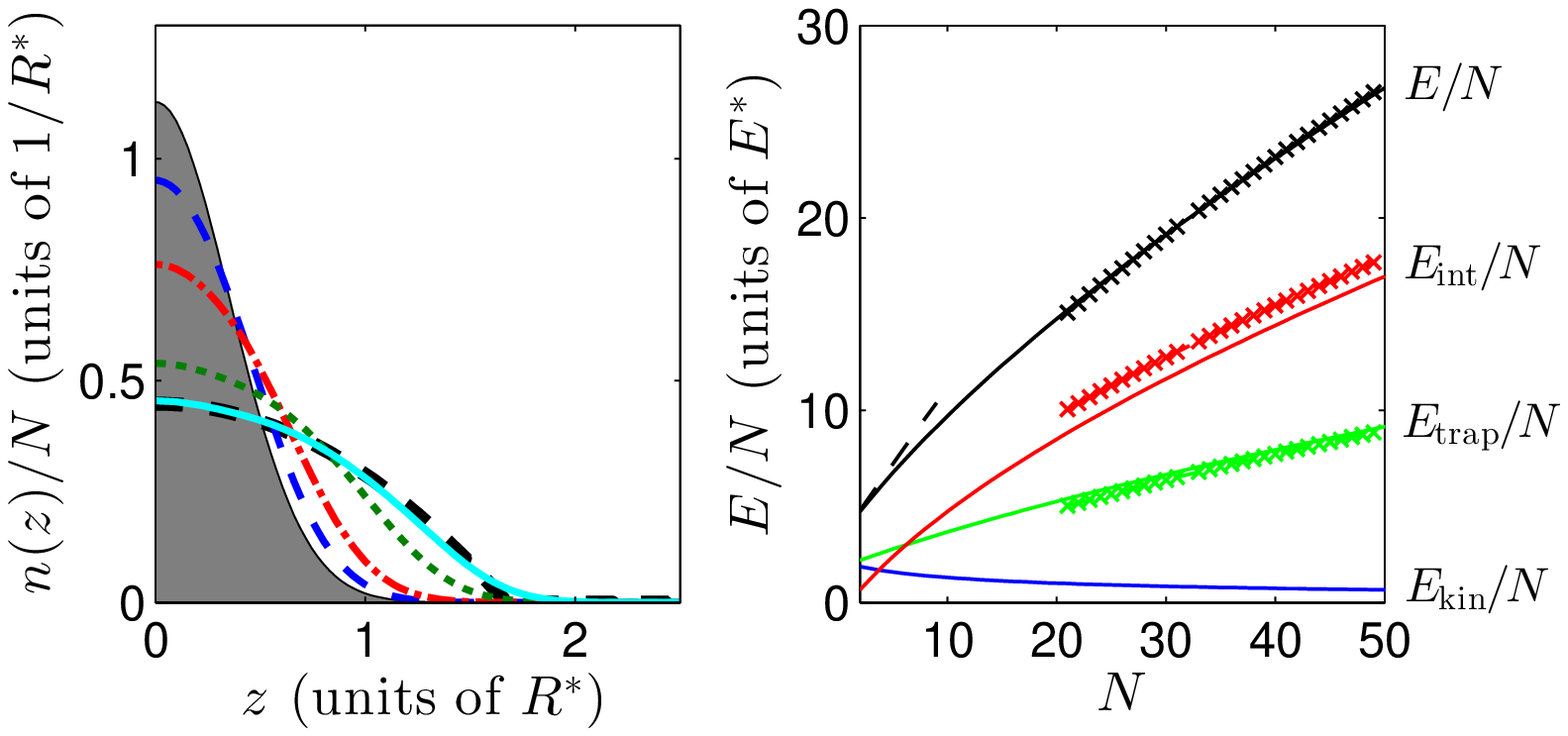}
\caption{(Color online) (Left) Atomic density distributions of the many-body ground state obtained
without ion for various values of the atomic number $N=4,10,30,50$ (dashed, dash-dotted, dotted, solid 
lines). 
The central peak height reduces with the atom number.
The gray shaded area represents the non interacting ground sstate $n_\mt{ideal}(z)$, whereas the thick black 
dashed line represents the
TF profile. The latter has been computed for $N=50$ atoms. Note that due to the symmetry of the
ground state it is sufficient to show the positive semi-axis.
(Right) Energy per particle and its  components.
The black, red, green, and blue lines represent the total, interaction, trapping, and kinetic
energy per particle, respectively. The perturbative solution of the total energy given in Eq.
\eqref{eq:GSGPlowNenergy} is displayed with the dashed line, whereas the TF results are indicated with the
crosses (x). Both approximations are only plotted in their range of validity.
In both panels, the interaction strength is $g = 2\,  E^* R^*$. 
}
\label{fig:HarmGS}
\end{figure}

\begin{figure}
 \includegraphics[width = 1.0\linewidth]{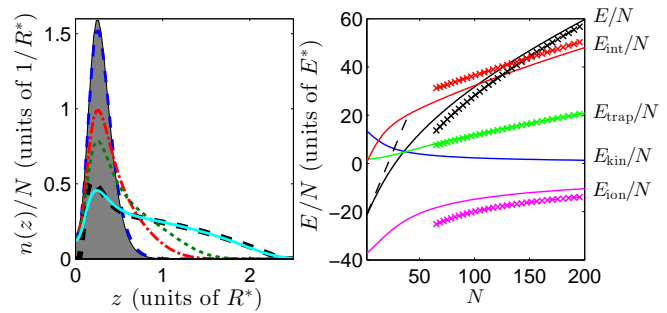}
\caption{(Color online) (Left) Atomic density distributions of the many-body ground state obtained with
ion for various values of the atomic number $N=4,30,50,150$ (dashed, dash-dotted, dotted, solid lines). 
The peak height reduces with the atom number.
The gray shaded area represents the
non interacting ground state $n_\mt{ideal}(z)$, whereas the thick black dashed line
represents the TF profile. The latter has been computed for
$N=150$ atoms. Note that due to the symmetry of the ground state it is sufficient to
show the positive semi-axis.
(Right) Energy per particle and its components (notation as in Fig. \ref{fig:HarmGS}). The
additional magenta line represents the ionic energy per particle. The
perturbative solution of the total energy given in Eq. \eqref{eq:GSGPlowNenergy} is displayed with the dashed
line, whereas the TF results are indicated with the crosses (x).  Both approximations are only plotted in
their range of validity. In both panels, the interaction strength is  $g = 2\,  E^* R^*$. }
\label{fig:IonGS}
\end{figure}

In the left panels of Figs. \ref{fig:HarmGS} and \ref{fig:IonGS}, the ground-state density $n(z)/N$ is 
plotted for several particle numbers for the harmonic and the ionic cases,
respectively. For small particle numbers, the
harmonic case reveals the expected Gaussian shape, whereas in the ionic case, we see a density hole at the
ion position and two peaks on each side. Atoms in these peaks are localized in the ion potential such that we
can think of them as being bound to the ion.
In both cases, the atomic density for $N=4$ can be well described by the non interacting
ground-state density distribution $n_\mt{ideal}(z)$ (see blue dashed line and gray shaded area). For the
ionic case this agreement looks even better.
Now, by increasing the atom number, one can observe a
broadening of the density distribution in both cases. The central region becomes depleted and the wings of the
distribution are populated, and therefore $n(z)\simeq n_\mt{ideal}(z)$ is not any longer a good
approximation. 
At large particle numbers, the density can be well described by the
TF approximation. Indeed, in the harmonic case, the density for $N=50$ (cyan solid line) rather
accurately reproduces the TF density $n_\mt{TF}(z)$ (thick black dashed line).
Also with the ion the TF distribution can be reproduced, even though larger particle numbers are needed. For 
$N=150$ (cyan solid line), the wings of the density distribution are in excellent agreement
with the TF profile (thick black dashed line), but close to the ion ($z=0$) deviations can be observed, in
particular for the two density peaks.

In the right panels of Figs. \ref{fig:HarmGS} and \ref{fig:IonGS}, we show the corresponding energy per
particle. In both cases, the total energy per particle (black solid
line) starts linearly for small atom numbers. This behavior is well captured by the perturbative
approximation of Eq. \eqref{eq:GSGPlowNenergy} (black dashed line).  However, the perturbative result quickly
deviates from the exact many-body calculation based on ML-MCTDHB as the number of atoms increases.
For larger particle numbers, the energy per particle bends over and reveals the expected $N^{2/3}$ behavior.
For the harmonic case, the TF approximations for the total energy per particle and its components are in good
agreement with the exact many-body simulations (crosses and solid curves), especially the total (black line)
and the trapping (green line) energy per particle match
the TF limit almost perfectly. On the other hand, the interaction energy component (red line) agrees less
well. The disagreement with the TF result can be easily understood by noting that the kinetic energy does not
vanish at $N=50$, and consequently the TF approximation is not optimally fulfilled.
Even though the ionic case reveals at first sight a similar behavior, we see that, apart
from the trapping energy, all energy components converge significantly slower to the TF solution, even
at rather large atom numbers. (Note that results for particle numbers up to $N=200$ are shown.)
Nevertheless, the TF curves can be nicely reproduced at such large particle numbers.

The fact that larger particle numbers compared to the harmonic situation are needed indicates the impact of
the ion on the transition from the ideal gas to the TF limit. In particular, it seems that the atom-ion
potential significantly ``slows'' this transition.

\subsubsection{Discussion of the emerging differences between the harmonic and the ionic case}

\begin{figure}
 \includegraphics[width = 1.0\linewidth]{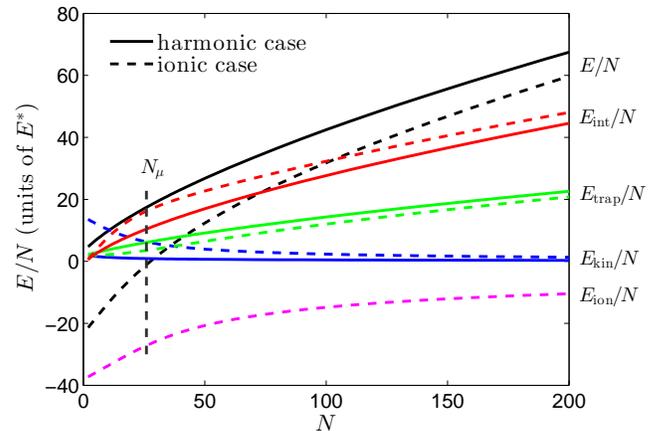}
\caption{ (Color online) Energy per particle $E/N$ (black) and its components $E_{\mt{kin}}$ (blue),
$E_{\mt{trap}}$ (green), $E_{\mt{int}}$ (red), and $E_\mt{ion}$ (magenta) for the harmonic (solid) and the
ionic (dashed) cases. The vertical line marks the particle number $N_\mu$ necessary for spatial
expansion.}
\label{fig:CompareHarmIonGS}
\end{figure}

In order to understand the differences between the harmonic and ionic case in more detail, we compare the
different components of the energy per particle for both cases in Fig. \ref{fig:CompareHarmIonGS}. 
Notably, the total energy per particle in the harmonic case (black solid line) is always above the one of the
ionic case (black dashed line) because of the additional (negative) ionic potential. 
They get closer for large $N$, but nevertheless a finite shift by $N_0$ particles between
$E_\mt{TF}^\mt{H}(N)/N$ and $E_\mt{TF}^\mt{I}(N)/N$ is present [see Eq. \eqref{eq:TFion}], which tends to
$N_0 \simeq 9$ in the limit $N\rightarrow\infty$ ($N \ge 380$).

Now for small particle numbers, we can see that the potential energy (green solid and dashed lines) for both
cases starts almost equally, meaning that the density has a comparable radius.
However, by enhancing the atom number, $E_{\mt{trap}}/N$ for the ionic case remains always below the one of
the harmonic case.  This indicates that the density distribution is more localized at $z=0$ (compare, e.g.,
 the left panels of Figs. \ref{fig:HarmGS} and \ref{fig:IonGS} for $N=30$). This is
also clearly visible in the interaction energy $E_{\mt{int}}/N$ in the presence of the ion. It grows more
rapidly than in the harmonic case for small particle numbers, as the particles are held together in the
ionic potential. Only at intermediate particle numbers ($N\sim30$), the particles start to
leak out of the ionic potential, leading to a relaxation of the interaction energy growth rate and allowing 
the $E_{\mt{trap}}/N$ to approach the trapping energy per particle of the harmonic case.
Furthermore, the energy per particle stemming from the ionic potential (magenta dashed line) quickly
rises at small
particle numbers. This shows that the density expands, but a small spatial spread needs a lot of energy per
boson due to the steep $-1/z^4$ potential. Only  when the bosonic density is able to leave the ionic
potential, the ionic part of the energy per particle flattens and crosses over to its TF behavior scaling
$N^{-1/3}$ [see also Eq. \eqref{eq:totalEion_TF}].

In order to better understand what actually impedes the transition to the TF regime and keeps the density from
broadening, we project the many-body ground state onto number states $|\vec{n}\rangle$ (see Sec.
\ref{secc:method} for the definition) of the non-interacting single particle basis $|\phi_i^0\rangle$. We find
that, apart from the state $|N,0,0,\cdots\rangle$, the number state
distribution of the ground state differs for the harmonic and the ionic cases. Indeed, without the
ionic potential, the number state which is the most populated due to interaction is the
$|N-1,0,1,\cdots\rangle$ state (note that $|N-1,1,0,\cdots\rangle$ is
forbidden due to symmetry). In contrast, in the ionic case the $|N-2,2,0,\cdots\rangle$ state is populated
the most. This difference can be understood by looking at the non-interacting single-particle energy
levels which are listed in Table \ref{tab:energs}. In the harmonic case,
we see the well-known equidistant energy spacing,
but for the ionic case pairs of energies exist which are close to each other.
Therefore, the state $|N-2,2,0,\cdots\rangle$ is energetically favorable in the presence of the ion.
Besides, since the two energetically lowest single-particle orbitals, corresponding to the even and the odd
bound states, are localized around the ion, the state $|N-2,2,0,\cdots\rangle$ does not contribute to
the spread of the wave function. Hence, the broadening of the density can only occur with energetically higher
number states (i.e., $n>1$).

\begin{table}
 \begin{ruledtabular}
  \begin{tabular}[c]{c||c|c|c|c|c}
     $E^{0}_n/E^*$ & $n=0$ & $n=1$ & $n=2$ & $n=3$ & $n=4$\\
    \hline
    $\mt{harm}$ 	 & 4.0 & 12.0& 20.0 & 28.0 & 36.0 \\
    $\mt{ion} $ 	 & -22.2 & -18.6 & 16.5 & 18.0 & 34.8 
  \end{tabular}
 \end{ruledtabular}
\caption{Harmonic and ionic single-particle energy $E^{0}_n/E^*$ of the first five single-particle states
$|\phi_n^0\rangle$.
\label{tab:energs}}
\end{table}

We can approximate the energy at which the broadening of the density becomes possible by using the third
single-particle energy level. By equating this energy with the chemical potential $\mu(N)=\partial E /\partial
N$, we can estimate the minimum particle number necessary for the expansion. By doing so we find $N_\mu = 26$,
which is marked by the vertical dashed black line in Fig. \ref{fig:CompareHarmIonGS}. It nicely
identifies the position where the interaction part of the energy per particle as well as the trapping energy
per particle for the ionic case change their slope.

In summary, we can interpret the observed differences between the ionic and the harmonic case by
thinking of the ion as a ``hole'' which needs to be filled before the growth in space and thus a behavior
comparable to the harmonic case can be observed.
This explains nicely the larger validity range of the perturbative results and the delayed crossover to
the TF regime.

\subsubsection{Momentum distribution and expansion}

\begin{figure}
 \includegraphics[width = 1.0\linewidth]{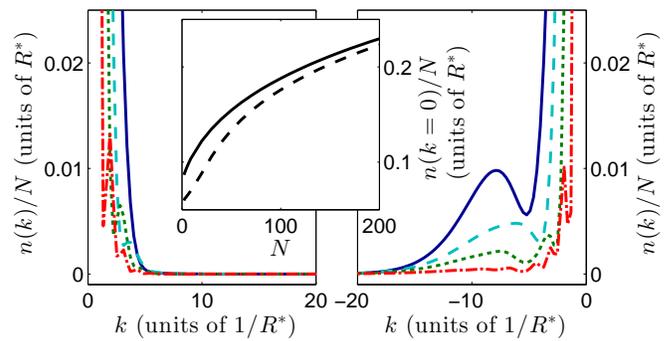}
\caption{ (Color online) The $k$-space density distribution for the harmonic (left panel) and the ionic case
(right panel) for
the particle numbers $N=2,20,50,200$ (solid, dashed, dotted, dash-dotted lines). The smaller the particle 
number is, the larger is the distribution width. Note that due to the symmetry of the $k$-space density 
distribution it
is sufficient to show one semiaxis. (Inset) Amplitude of the momentum distribution $n(k)$ at $k=0$
as a function of the particle number $N$ for the harmonic (solid line) and the ionic (dashed line) case. }
\label{fig:kspaceDis}
\end{figure}

Let us next investigate the (experimental accessible) momentum distribution $n(k)$,
\begin{equation}\label{eq:FT}
 n(k) = \frac{1}{2\pi}\iint \d{z}\d{z^\prime} \rho_1(z,z^\prime) e^{-ik(z-z^\prime)}.
\end{equation}
It can be used as a measure for coherence, since it incorporates the on- and off-diagonal contributions of
the one-body reduced density matrix. We can start the analysis by looking at the amplitude of the peak at zero
momentum $n(k=0)$, recalling that for a homogeneous non interacting Bose condensate at zero temperature $n(k)$
is a $\delta$ function at $k=0$.
The inset in Fig. \ref{fig:kspaceDis} shows the value $n(k=0)$ as a function of the particle number with and
without the ion. Due to the presence of the ion, the coherence is smaller for all particle numbers studied
here, since the induced density hole reduces the coherence to the central region. Only for high particle
numbers, both cases become comparable, as we have already seen in the previous section in terms of
energy and spatial density. The reduced peak height of the momentum distribution goes hand in hand with its
broadening. Therefore, we show in Fig. \ref{fig:kspaceDis} the wings of the $k$-space distribution for the
harmonic case (left panel) and the ionic case (right panel) for multiple particle numbers. In the harmonic 
case,
the distribution is Gaussian for $N=2$ and shrinks in width for growing $N$.
In contrast, the ionic case is much broader due to two side peaks at $k \approx \pm 10/R^* $ for small
particle numbers. Nevertheless, for high particle numbers, when the system enters the TF regime, the
distribution has sharpened again and the side peaks have vanished. For $N=200$, both cases exhibit a
similar momentum distribution.
We can explain the broader distribution for the ionic case by the overall loss of coherence due to the
separation of the cloud into two parts by the ion.  The bimodal structure at the neck of the
momentum distribution, however, arises due to the coherence of particles in the left and the right peaks of
the spatial density distribution. Their distance dictates the positions of the peaks in the momentum
distribution.

Since the momentum distribution for the harmonic case is approximatively zero at the values $k \approx \pm
10/R^* $ one could use these two side peaks as an indicator for the presence of an ion in a experiment, e.g.,
by time-of-flight measurements. Nevertheless, such a signal might be hard to detect due to the small amplitude
of the peaks.

A further possibility to check the presence of the ion could be to measure the atomic density during the
expansion in a quasi-1D waveguide, as experimentally realized by Bongs \textit{et al.} \cite{Bongs2001}.
In order to simulate such an expansion experiment, we study the temporal evolution of the interacting
atomic cloud after the removal of both potential terms, namely the trapping and the atom-ion potentials.
The former can be switched off in a time much smaller than the inverse of the trap frequency, whereas
removing the latter might have an impact on the atomic cloud due to the long-range nature of the atom-ion
interaction. Here, however, we neglect these effects and simply switch off the atom-ion interaction.
Additionally, we note that during the expansion the atomic quantum gas is not any longer in the weakly
interacting regime, since the Lieb-Liniger parameter scales as $\gamma_{LL} \approx 1/\rho$
with the mean density $\rho$, and therefore the interaction becomes larger and larger.
Because of this anomalous behavior in 1D, the total expansion time we can actually simulate is limited by
the number of used SPFs.

\begin{figure*}
  \includegraphics[width = 1.0\linewidth]{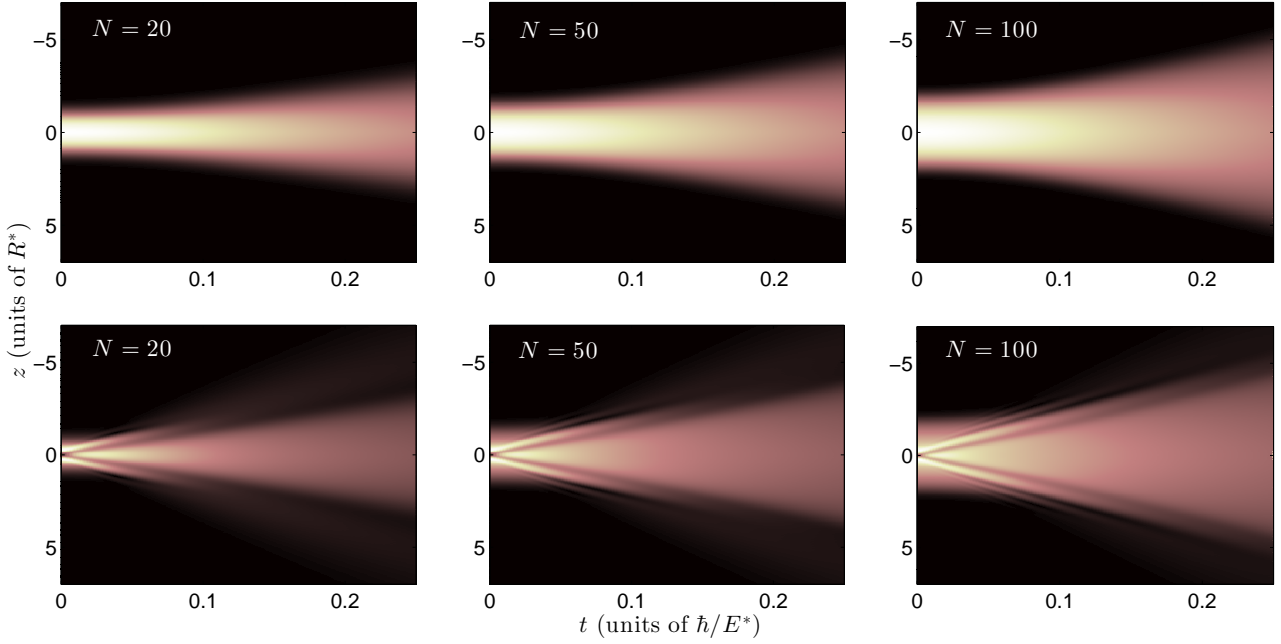}
 \caption{ (Color online) Density profile $n(z,t)/N$ of the atomic cloud during an expansion. The expansion
starts from the  ground state of the harmonic case (top row) and the ionic (bottom row) case for $N=20, 50, 
100$. The time $t$ is given in units of $\hbar/E^*$. For a $^{87}$Rb atom we have $\hbar/E^* \approx 0.38\,
\milli\second$, whereas for $^{7}$Li it corresponds to $0.0013\,\milli\second$.}
 \label{fig:expandHvsI}
 \end{figure*}

In Fig. \ref{fig:expandHvsI}, we show the density $n(z)/N$ during the expansion. At $t=0$, the system is in
the ground state and for $t>0$ both the trap and the atom-ion interaction are switched off.
In the harmonic case, the atomic cloud just expands in space as a single macroscopic object for all
particle numbers as it can be seen in Fig. \ref{fig:expandHvsI} (top row). In contrast, the time evolution
of the ionic case, Fig. \ref{fig:expandHvsI} (bottom row), shows clear interference fringes. These stem
from the interference of ``particles'' from the left and the right sides of the ion. They show the 
interference pattern at $|z| > 0$ due to the different path lengths they have traveled during the expansion.

In conclusion, one can use the reduced peak amplitude $n(k=0)$, the bimodal structure in the momentum
distribution, or the interference patterns from the 1D expansion to prove the presence
of the ion within the
atomic cloud. 

\subsection{Strong interactions}

Now, we leave the mean-field regime and face the situation of an interaction dominated system. This
means that the interaction becomes strong enough to deplete significantly the first natural orbital making a
multi-orbital description of the system essential \cite{Alon2005}. We want to study this fragmentation process
up to the
fermionization limit characterized by $g \rightarrow \infty$. In this limit, the
Bose-Fermi mapping can be applied. The strong repulsive forces deplete the interaction regions $\{z_i=z_j \}$
for particles $i$ and $j$, which emulates the Pauli exclusion principle for fermions.
Thus, our strongly interacting bosonic system, also called Tonks-Girardeau (TG) gas, behaves like a
non-interacting fermionic system \cite{Girardeau1960}. In this case, for example, the density $n_\mt{TG}(z) =
\sum_{i=0}^{N-1}|\phi_i^0(z)|^2 $ and the total energy $E_\mt{TG} = \sum_{i=0}^{N-1} E_i^0$ can be determined
analytically.
With our method, however, we can study the transition for increasing $g$ to the TG limit. It was shown
that due to the used truncated Hilbert space, introduced by the ML-MCTDHB, this limit can be achieved for a
finite interaction strength $g_0$ \cite{Ernst2011}. Thus, we can compare our simulations to these exact
expressions even though a finite $g$ is used. Moreover, a rescaling procedure makes it possible to identify 
the physical interaction strength $g_\mt{phys}$, leading to solutions which are independent of the chosen 
Hilbert space truncation.

The harmonic case has been extensively studied from the condensation via the fragmentation up to the
fermionization limit \cite{Dunjko2001,Alon2005}. Here we only summarize the most
relevant results which will be used in our later analysis.
In the fermionization limit, the lowest natural population scales like $n_0/N \propto
N^{-0.41}$ in a harmonic trap \cite{Girardeau2001}, which is in contrast to the homogeneous case ($N^{-0.5}$)
and to the fermionic system in a harmonic trap ($N^{-1}$).
A distinct feature is the transition of the density profile $n(z)$ from the Gaussian to
a multi hump structure revealing as many peaks as particles in the system \cite{Kolomeisky2000}.  Furthermore,
the momentum distribution, still having a strong peak at $k=0$, spreads to higher $k$ values, revealing a
universal decay $\propto k^{-4}$ in the asymptotic regime \cite{Minguzzi2002} due to the short-range contact
interaction. Finally, exemplary one- and two-body reduced density matrices for the harmonic case can be
found in Refs. \cite{Zollner2006a} and \cite{Zollner2006}, respectively.

\begin{figure}
 \includegraphics[width = 1.0\linewidth]{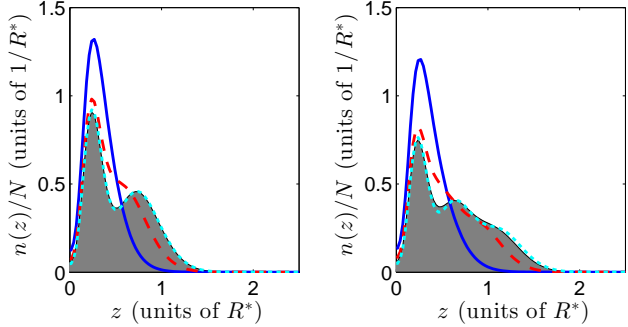}
 % phi_20.eps: 0x0 pixel, 300dpi, 0.00x0.00 cm, bb=   18   179   594   612
 \caption{(Color online) Bosonic density for large values of the interaction strength $g = {10,40,160}
\,E^*R^*$ (solid, dashed, dotted lines) for $N=4$ (left panel) and $N=5$ (right panel) using $m=10$
orbitals with the ion located in the
center of the harmonic confinement. The peak height reduces with the interaction strength. Additionally,
we plot the Tonks-Girardeau density distribution
$n_\mt{TG}(z)/N$ as a gray shaded area. Note that due to the symmetry of
the ground state it is sufficient to show the positive semiaxis.}
 \label{fig:ionFermg}
\end{figure}

Now let us analyze the situation in the presence of an ion. In Fig. \ref{fig:ionFermg} (left
panel), the density distribution for $N=4$ for the ionic case is plotted for different interaction strengths
using $m=10$ orbitals. We observe the development of two extra side peaks with respect to the non-interacting
ground state.
For $g=160\,R^*E^*$, the result is perfectly consistent with the fermionized density $n_\mt{TG}(z)$ shown as
gray shaded area. Thus, the fermionized density for the ionic case with $N=4$
atoms also reveals an $N$-hump
structure which has been conjectured to be a universal feature for a general trapping potential
\cite{Alon2005}.
Nevertheless, for an odd number of particles the density cannot have an odd number of peaks. For symmetry
reasons, this would require a central peak at $z=0$, but this is hindered by the
presence of the ionic potential. 

In Fig. \ref{fig:ionFermg} (right panel), we show the evolution of the
density profile for increasing interaction strength $g$ for $N=5$ bosons using $m=10$ orbitals. The TG density
profile is nicely reproduced for high $g$. Further, we can clearly see that the additional particle is
distributed mostly in the outer region of the density profile leading to shoulders on both sides of the
distribution. Therefore, the two additional side peaks are slightly pushed inwards compared to the $N=4$
case (see the left panel of Fig. \ref{fig:ionFermg})  due to the repulsive interaction between the atoms. 

\begin{figure}
 \includegraphics[width = 1.0\linewidth]{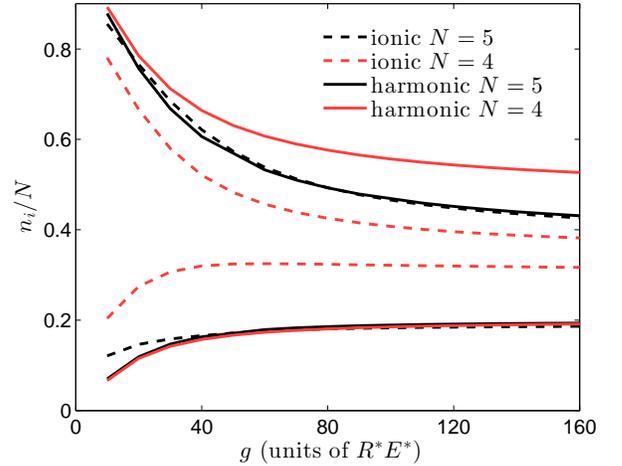}
 % phi_20.eps: 0x0 pixel, 300dpi, 0.00x0.00 cm, bb=   18   179   594   612
 \caption{(Color online) Natural populations $n_0/N$ (four upper lines) and $n_1/N$ (four lower
lines) for the harmonic (solid lines) and the ionic
(dashed lines) case for $N=4$ (red lines) and $N=5$ (black lines) particles in dependence of the interaction
strength $g$. Note that even though $n_i$ for $i>1$ are not plotted here they are non-zero.}
 \label{fig:npop}
\end{figure}

In order to better understand this behavior, we have investigated the evolution of the populations of the
natural orbitals for increasing $g$. Figure \ref{fig:npop} shows the natural populations $n_0/N$ and $n_1/N$,
introduced in Eq. \eqref{eq:natpopOrb}, for an even ($N=4$, red) and
an odd ($N=5$, black) particle number for the ionic (dashed lines) as well as for the harmonic (solid lines)
case. For the latter, the fragmentation is stronger for the case of $N=5$ particles, since $n_0/N \sim
N^{-0.41}$. Interestingly, this behavior is opposite for the ionic case. For $N=4$ particles, we can see
a significantly stronger fragmentation for all interaction strengths. Further, $n_1/N$ becomes comparable
to $n_0/N$
in this case. The former can be explained by noting that the ion separates the cloud into two
parts leading to two degenerate subsystems with $N/2$ atoms. Thus, for an even number of atoms, the ion
assists the fragmentation process and finally enhances the fragmentation in the fermionization limit.
However, for $N=5$ we can observe nearly the same behavior for all $g$ as in the harmonic case with the
same atom number.
This behavior can be understood as follows: The additional particle,
which is distributed on both sides of the cloud, as we have seen above, counteracts the fragmentation and
preserves the spatial coherence of the system. On the other hand, the density hole induced by
the ion is also present for an odd particle number and should enhance the fragmentation comparably. These two
competing processes seem to nearly balance each other, leading only to a slight enhancement of fragmentation
compared to the harmonic case.

\begin{figure}
 \includegraphics[width = 1.0\linewidth]{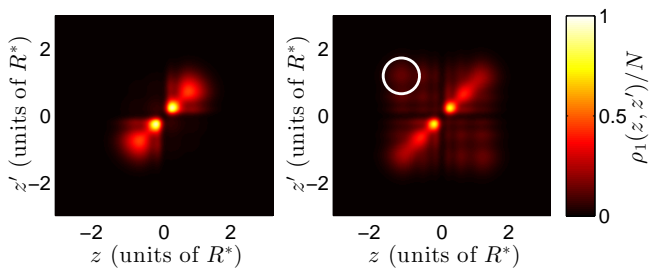}
 % phi_20.eps: 0x0 pixel, 300dpi, 0.00x0.00 cm, bb=   18   179   594   612
 \caption{(Color online) Reduced one-body density matrix for the ionic case for $N=4$ (left panel) and $N=5$
(right panel). The white circle in the right panel indicates the region of strong off-diagonal
coherence.}
 \label{fig:OneBDM}
\end{figure}

These conclusions are further supported by looking at the one-body reduced density matrices.
In Fig. \ref{fig:OneBDM}, we plot $\rho_1(z,z^\prime)$ [see Eq. \eqref{eq:natpopOrb} for its definition] for
$N=4$ (left panel)  and $N=5$ (right panel) atoms in the limit $g \rightarrow \infty$. For $N=4$, we can
clearly see that the two parts of the atomic cloud have lost their
coherence completely. Our numerical finding is also supported by a recent study  \cite{Lelas2009}, where it
has been analytically proven that for a TG with an even number of particles and with an infinite central
barrier off-diagonal correlations are negligible. This explains the enhanced
fragmentation induced by the presence of the ion.
In contrast, the $N=5$ case exhibits still significant off-diagonal contributions which are most prominent
between the two outermost density accumulations at $|z|\sim 1.5 R^*$, as indicated by the
white circle in Fig. \ref{fig:OneBDM}. These strong coherences almost perfectly compensate the loss of
coherence around $z,z^\prime = 0$ due to the ionic potential.

\begin{figure}
 \includegraphics[width = 1.0\linewidth]{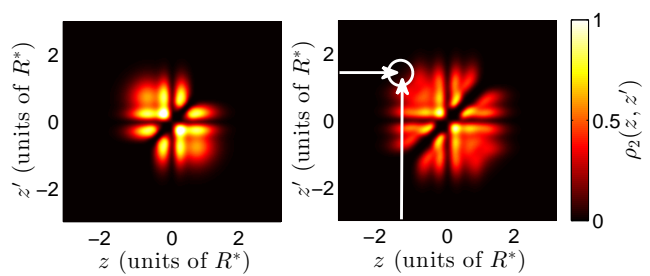}
 % phi_20.eps: 0x0 pixel, 300dpi, 0.00x0.00 cm, bb=   18   179   594   612
 \caption{(Color online) Reduced two-body density matrix for the ionic case for $N=4$ (left panel)  and $N=5$
(right panel).
The white circle in the right panel indicates the strong suppression of correlation between
two particles at the two positions marked by white arrows.}
 \label{fig:TwoBDM}
\end{figure}

We complete the analysis of the fragmentation process by inspecting the diagonal of the reduced two-body
density matrix $\rho_2(z,z')$ defined by Eq. \eqref{eq:g2}. In Fig.
\ref{fig:TwoBDM}, $\rho_2(z,z')$ is shown in the fermionization limit again for $N=4$ and $N=5$. In the case
of an even particle number, we observe the characteristic depletion of the diagonal as well as the
 ``checkerboard'' pattern known from the harmonic case. The effect of the ionic potential can only be observed
in the strong suppression of correlation at $z,z'=0$. We can understand the distribution in the following way:
Imagine that we have one particle in one of the four peaks of the density distribution $n(z)$ (see the left
panel of Fig. \ref{fig:ionFermg}). Then the
probability to find another particle in each of the other three peaks is nearly one. 
The situation is different for $N=5$. For a
particle in one of the four peaks (see the right panel of Fig. \ref{fig:ionFermg}), we can
observe enhanced
probability to find a second particle at the position of the three other peaks and
additionally at the shoulders of the distribution.
Since there are only four other particles, one particle has to be delocalized.
In addition to this, no correlation is found between the left (vertical arrow) and right (horizontal
arrow)  density shoulders at  $z \sim -1.5R^*$ and $z^\prime \sim 1.5R^*$,
respectively, which we highlight by the white circle in Fig. \ref{fig:TwoBDM}.
This indicates that these
additional shoulders do not result from two different particles.  Therefore, the picture of having one
particle distributed on both sides of the atomic cloud is qualitatively valid.

Finally, it is interesting to compare the behavior of the ionic case with two other systems that have been
discussed in the literature. First, we can look at a double-well setup investigated by Z{\"o}llner \textit{et 
al.} \cite{Zollner2006a,Zollner2006}.
They found by successively enlarging $g$ that an even particle number assists while an odd particle number
delays the fragmentation process compared to the harmonic case. Nevertheless, this difference vanishes for
increasing $g$ when the system has enough interaction energy to start filling the central hole induced by the
potential with the extra particle. 
In the ionic case, the ionic potential forces the atomic wavefunction to be zero near the origin [see
Eqs. \eqref{eq:psiEpaper} and \eqref{eq:psiOpaper}] which is achieved by using a large Gaussian height
$\mt{v}_0$.
Therefore, the density hole is not  filled even for infinite values of $g$ or higher particle numbers.
Instead, a density bubble forms in the atomic cloud \cite{Goold2010}. Note that the density hole observed here
is smaller than the one found in Ref. \cite{Goold2010} since we included  the two highest bound states.
Second, we can compare the ionic case to a so-called split trap, a harmonic trap with a superimposed
central $\delta$ peak $\kappa \delta(z)$, as it has been recently investigated in Refs. 
\cite{Busch2003,Goold2008,Goold2010a}.
Even for large $\kappa$, it has been found that the density hole is preserved. Further, the behavior of the
natural populations is comparable to the one of the ionic case reported here.

Hence, we can conclude that an atomic cloud with an immersed static ion shows features which are comparable to
the ones of a double-well setup. Nevertheless, this analogy fails in the fermionization limit where some
properties of the system can better be compared to a split trap, as already pointed out by Goold \textit{et 
al.} \cite{Goold2010}.

\subsubsection{Expansion}

Fragmentation reduces the ability to interfere. Therefore, the above-discussed difference in the fragmentation
originating from particle number parity can be experimentally observed within an expansion measurement.
As before, we remove the harmonic and the ionic potentials and propagate the above-obtained ground states in
time in order to simulate the expansion in a quasi-1D waveguide.

\begin{figure*}
 \includegraphics[width = 1.0\linewidth]{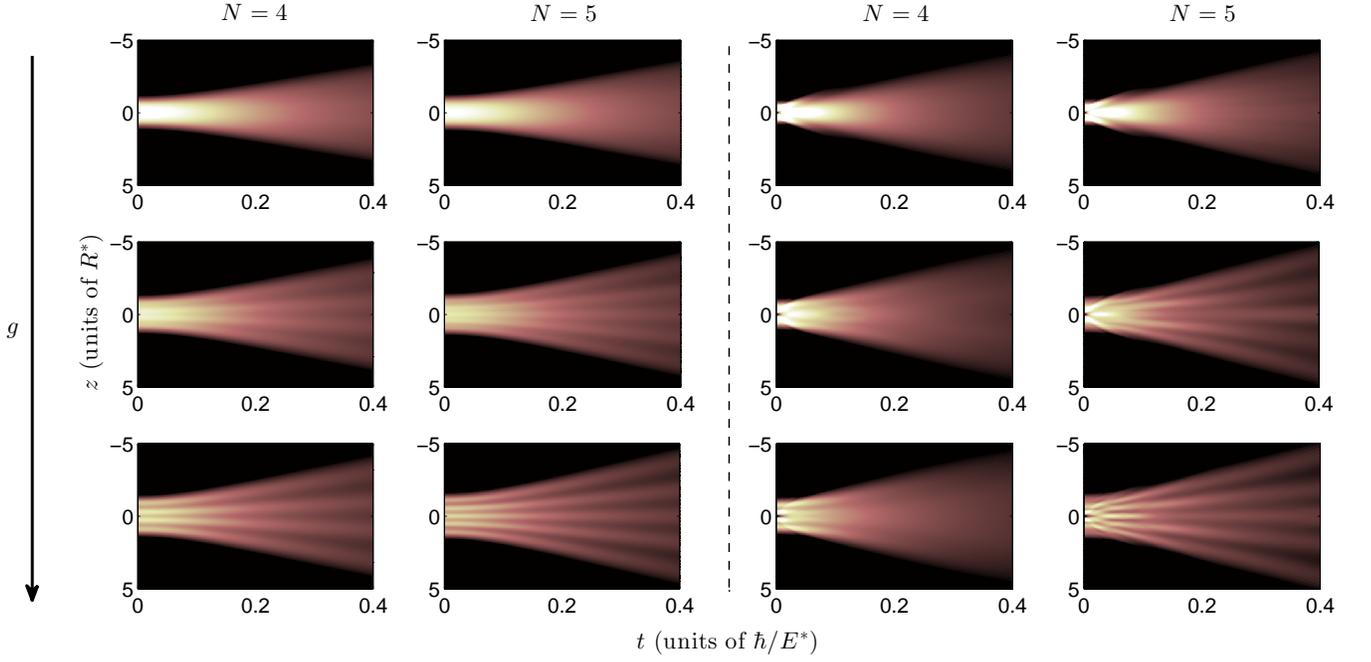}
 % phi_20.eps: 0x0 pixel, 300dpi, 0.00x0.00 cm, bb=   18   179   594   612
 \caption{ (Color online) Density profile $n(z,t)/N$ of the atomic cloud during an expansion for increasing
$g$. The expansion starts from the  ground state of the harmonic case with $N=4$ (first column) and $N=5$
(second column) and of the ionic case with $N=4$ (third column) and $N=5$ (fourth column). Results are shown
for $g=10$ (top row), $g=30$ (middle row), and $g=160$ (bottom row)  in units of $E^*R^*$. Typical
time scales are given in the caption of Fig. \ref{fig:expandHvsI}}
 \label{fig:expansionStrong}
\end{figure*}

In Fig. \ref{fig:expansionStrong}, we show the time-dependent density profile for the harmonic and the ionic
cases for $N=4$ and $N=5$ particles. Starting with the harmonic case (first and second columns), we can see 
that for $g=10\, E^* R^*$ (top row) the expansion shows for $N=4,5$ a similar behavior as in the weakly 
interacting case (compare Fig. \ref{fig:expandHvsI}).
At intermediate $g=30\, E^* R^*$ (middle row), density modulations are visible in the expansion which reflect
the $N$-hump structure of the initial profile which starts to appear as we approach the TG regime. 
In the limit $g\rightarrow \infty$ (bottom row), the $N$-hump structure becomes clearly visible and is
preserved during the expansion. 
In the ionic case (third  and fourth columns), the behavior for $N=4,5$ and $g=10\, E^* R^*$ (top row) is 
also similar to that of the weakly
interaction regime (see Fig. \ref{fig:expandHvsI}). Nevertheless, for an increased interaction strength of
$g=30\, E^* R^*$ (middle row), we can observe that an even ($N=4$) particle number and an odd ($N=5$) 
particle number lead to
different expansion behaviors. For $N=4$, a rather flat density profile shows up during the expansion while
for $N=5$ five density peaks appear. This difference becomes even more pronounced for $g\rightarrow \infty$
(bottom row). We can understand this expansion behavior as follows. After the removal of the harmonic and
ionic potentials, the left and the right parts of the density distribution (see Fig. \ref{fig:ionFermg}) start
to penetrate into each other. In the case of four atoms, both sides are completely incoherent, as we have seen
in Fig. \ref{fig:OneBDM} (left panel), and therefore no structure appears during the expansion. In contrast,
the additional particle for $N=5$ establishes coherence between the two sides (see the right panel of Fig.
\ref{fig:OneBDM}) such that interference fringes occur during the expansion. We note that a similar behavior
has been observed for a TG gas with a Dirac’s $\delta$ in the center of the trap \cite{Goold2008}. 
This signature can be used to distinguish between an even atom number and an odd atom number. Hence, we can 
conclude that the atom number parity in the presence of the ion can be measured by a quasi-1D expansion. 
 
%%% End Section IV
%%%%%%%%%%%%%%%%%%%%%%%%%%%%%%%%%%%%%%%%%%%%%%%%%%%%%%%%%%%%%%%%%%%%%%%%%%%%%%%%%%%%%%%%
%%% Section V

\section{Conclusions and outlooks}

In this paper, we have presented many-body calculations concerning the ground-state properties of
the hybrid multi-atom single-ion system. To accomplish this task, we have first introduced a model potential
for the two-body atom-ion interaction which is able to reproduce the QDT results. This was
an essential ingredient for the subsequent investigation via the ML-MCTDHB. We have then investigated in 
detail the transition from the
weak to the strong interaction regime for a single ion immersed in a bosonic atomic cloud. 
For weakly interacting bosons, we found, by increasing the atom number $N$, that the ion impedes the
transition from the ideal gas behavior to the TF limit. We showed that this effect can be exploited
in expansion experiments in order to prove the presence of an ion.
On the other hand, in the strong interaction regime, we observed that the ion assists the
fragmentation process and enhances the fragmentation in the fermionization limit for an even atom number. In
contrast, the fragmentation is nearly unaffected by the ion for an odd $N$. We explained  this
behavior by the spatial splitting of the additional particle which counteracts the fragmentation. Further, we
showed that this difference for even and odd particle numbers can be observed by looking at the
interference in an expansion experiment.

Note that in view of the experimental verification of our results, one has to face the problem of atom
loss from the trap which we neglected in the present work. Loss can result from three-body scattering
which might become important in the weakly interacting regime \cite{Gangardt2003}, from spontaneous scattering
induced by high power traps with decay times of several hundred milliseconds \cite{Serwane2011}, and from
atom-ion collisions which can become the dominant loss channel if the ion is not cooled to the ultracold
regime \cite{Schmid2010a}.

In conclusion, the present results can be viewed as the first step towards the simulations of such hybrid
quantum many-body system. Since our method is specifically designed for the simulation of multi species
systems, an obvious extension of our work is the study of the ground state when the ion motion is included,
which will be pursued in the near future. Moreover, this work opens the way for the study of dynamical
processes within the atom-ion hybrid system.
Focusing on the atomic cloud, one can think of the investigation of the dynamical formation of a density
bubble  and charged molecules within the atomic cloud, and of tunneling dynamics, e.g., in a bosonic Josephson
junction setup \cite{Gerritsma2012}. With respect to the ion, the study of sympathetic cooling, i.e., energy
transfer from an excited ion to the atomic cloud, and the impact of micromotion are also of current interest
and could be possibly simulated with our technique. Even the inclusion of spin degrees of freedom is
possible in our method allowing for the study of a single-ion qubit in an atomic bath as recently reported in
Ref. \cite{Ratschbacher2013}. Finally, one could extend the setup to multiple ions making it possible, for 
instance, to investigate the polaron physics emerging in such hybrid systems.

%%% End Section IV
%%%%%%%%%%%%%%%%%%%%%%%%%%%%%%%%%%%%%%%%%%%%%%%%%%%%%%%%%%%%%%%%%%%%%%%%%%%%%%%%%%%%%%%%
%%% Acknowledgments

\begin{acknowledgments}
This work has been supported by the excellence cluster 'The Hamburg Centre for
Ultrafast Imaging - Structure, Dynamics and Control of Matter at the Atomic Scale' of
the Deutsche Forschungsgemeinschaft. A. N. acknowledges discussions with Rene Gerritsma and J. S. with Sven
Kr\"onke and Lushuai Cao concerning ML-MCTDHB.
\end{acknowledgments}

%%% End Acknowledgments
%%%%%%%%%%%%%%%%%%%%%%%%%%%%%%%%%%%%%%%%%%%%%%%%%%%%%%%%%%%%%%%%%%%%%%%%%%%%%%%%%%%%%%%%
%%% Appendix

\appendix*

\section{Thomas-Fermi limit for the ionic case}\label{app:TFIon}

The energy per particle in the TF regime cannot be derived analytically for the ionic case.
Nevertheless, we can find approximative expressions for large particle numbers. In this limit, the outer TF
radius $z_\mu^O$ is large enough that we can neglect the ionic potential at the borders of the density
distribution. Therefore, integrals over the ionic potential [see Eq. \eqref{eq:GSTFenergy}] become independent
of the particle number $N$, making it possible to identify the chemical potential via the normalization of the
wave function,
\begin{equation}
 \mu(N) = \left[ (N-N_0)\frac{3g}{4 l_\parallel^2} \right]^{2/3}.
\end{equation}
This leads to an energy per particle of
\begin{equation}\label{eq:APPTFion}
 \frac{E_\mt{TF}^\mt{I}(N)}{N}  = \frac{3 }{5} \frac{e_0(N-N_0)^{5/3}}{N} + \frac{E^\mt{I}_0}{N}.
\end{equation}
In the same way, we can approximate the single components of the energy per particle by
\begin{align}
   \frac{E_\mt{TF,int}^\mt{I}(N)}{N} &=  \frac{2 }{5} \frac{e_0(N-N_0)^{5/3}}{N} \nonumber \\
&+ N_0 e_0 \frac{(N-N_0)^{2/3}}{N} - \frac{E^\mt{I}_0}{N}, \label{eq:totalEww_TF}\\
   \frac{E_\mt{TF,trap}^\mt{I}(N)}{N} &=  \frac{1 }{5} \frac{e_0(N-N_0)^{5/3}}{N}  + \frac{E^\mt{I}_1}{N}, 
\label{eq:totalEtrap_TF} \\
    \frac{E_\mt{TF,ion}^\mt{I}(N)}{N} &= N_0 e_0 \frac{(N-N_0)^{2/3}}{N} + \frac{E^\mt{I}_2}{N},
\label{eq:totalEion_TF}
\end{align}
with the constans
\begin{align}
  E^\mt{I}_1 &= -1/g\int_{-z_\mu^O}^{z_\mu^O} [ V_\mt{trap}(z)V_\mt{mod}(z) ] \d{z},  \\
  E^\mt{I}_2 &= -1/g\int_{-z_\mu^O}^{z_\mu^O} [ V_\mt{trap}(z)V_\mt{mod}(z) + V_\mt{mod}(z)^2] \d{z} ,
\end{align}
which have the property $ E^\mt{I}_1 + E^\mt{I}_2 = 2E^\mt{I}_0$.

%%% End Appendix
%%%%%%%%%%%%%%%%%%%%%%%%%%%%%%%%%%%%%%%%%%%%%%%%%%%%%%%%%%%%%%%%%%%%%%%%%%%%%%%%%%%%%%%%
%%% Bibliography

\bibliography{library}

\begin{thebibliography}{75}%
\makeatletter
\providecommand \@ifxundefined [1]{%
 \@ifx{#1\undefined}
}%
\providecommand \@ifnum [1]{%
 \ifnum #1\expandafter \@firstoftwo
 \else \expandafter \@secondoftwo
 \fi
}%
\providecommand \@ifx [1]{%
 \ifx #1\expandafter \@firstoftwo
 \else \expandafter \@secondoftwo
 \fi
}%
\providecommand \natexlab [1]{#1}%
\providecommand \enquote  [1]{``#1''}%
\providecommand \bibnamefont  [1]{#1}%
\providecommand \bibfnamefont [1]{#1}%
\providecommand \citenamefont [1]{#1}%
\providecommand \href@noop [0]{\@secondoftwo}%
\providecommand \href [0]{\begingroup \@sanitize@url \@href}%
\providecommand \@href[1]{\@@startlink{#1}\@@href}%
\providecommand \@@href[1]{\endgroup#1\@@endlink}%
\providecommand \@sanitize@url [0]{\catcode `\\12\catcode `\$12\catcode
  `\&12\catcode `\#12\catcode `\^12\catcode `\_12\catcode `\%12\relax}%
\providecommand \@@startlink[1]{}%
\providecommand \@@endlink[0]{}%
\providecommand \url  [0]{\begingroup\@sanitize@url \@url }%
\providecommand \@url [1]{\endgroup\@href {#1}{\urlprefix }}%
\providecommand \urlprefix  [0]{URL }%
\providecommand \Eprint [0]{\href }%
\providecommand \doibase [0]{http://dx.doi.org/}%
\providecommand \selectlanguage [0]{\@gobble}%
\providecommand \bibinfo  [0]{\@secondoftwo}%
\providecommand \bibfield  [0]{\@secondoftwo}%
\providecommand \translation [1]{[#1]}%
\providecommand \BibitemOpen [0]{}%
\providecommand \bibitemStop [0]{}%
\providecommand \bibitemNoStop [0]{.\EOS\space}%
\providecommand \EOS [0]{\spacefactor3000\relax}%
\providecommand \BibitemShut  [1]{\csname bibitem#1\endcsname}%
\let\auto@bib@innerbib\@empty
%</preamble>
\bibitem [{\citenamefont {H\"{a}rter}\ and\ \citenamefont {{Hecker
  Denschlag}}(2014)}]{Harter2014}%
  \BibitemOpen
  \bibfield  {author} {\bibinfo {author} {\bibfnamefont {A.}~\bibnamefont
  {H\"{a}rter}}\ and\ \bibinfo {author} {\bibfnamefont {J.}~\bibnamefont
  {{Hecker Denschlag}}},\ }\href {\doibase 10.1080/00107514.2013.854618}
  {\bibfield  {journal} {\bibinfo  {journal} {Contemp. Phys.}\ }\textbf
  {\bibinfo {volume} {55}},\ \bibinfo {pages} {33} (\bibinfo {year}
  {2014})}\BibitemShut {NoStop}%
\bibitem [{\citenamefont {Cetina}\ \emph {et~al.}(2012)\citenamefont {Cetina},
  \citenamefont {Grier},\ and\ \citenamefont {Vuleti\'{c}}}]{Cetina2012}%
  \BibitemOpen
  \bibfield  {author} {\bibinfo {author} {\bibfnamefont {M.}~\bibnamefont
  {Cetina}}, \bibinfo {author} {\bibfnamefont {A.~T.}\ \bibnamefont {Grier}}, \
  and\ \bibinfo {author} {\bibfnamefont {V.}~\bibnamefont {Vuleti\'{c}}},\
  }\href {\doibase 10.1103/PhysRevLett.109.253201} {\bibfield  {journal}
  {\bibinfo  {journal} {Phys. Rev. Lett.}\ }\textbf {\bibinfo {volume} {109}},\
  \bibinfo {pages} {253201} (\bibinfo {year} {2012})}\BibitemShut {NoStop}%
\bibitem [{\citenamefont {Krych}\ and\ \citenamefont
  {Idziaszek}(2015)}]{Krych2015}%
  \BibitemOpen
  \bibfield  {author} {\bibinfo {author} {\bibfnamefont {M.}~\bibnamefont
  {Krych}}\ and\ \bibinfo {author} {\bibfnamefont {Z.}~\bibnamefont
  {Idziaszek}},\ }\href {\doibase 10.1103/PhysRevA.91.023430} {\bibfield
  {journal} {\bibinfo  {journal} {Phys. Rev. A}\ }\textbf {\bibinfo {volume}
  {91}},\ \bibinfo {pages} {023430} (\bibinfo {year} {2015})}\BibitemShut
  {NoStop}%
\bibitem [{\citenamefont {Schneider}\ \emph {et~al.}(2010)\citenamefont
  {Schneider}, \citenamefont {Enderlein}, \citenamefont {Huber},\ and\
  \citenamefont {Schaetz}}]{Schneider2010}%
  \BibitemOpen
  \bibfield  {author} {\bibinfo {author} {\bibfnamefont {C.}~\bibnamefont
  {Schneider}}, \bibinfo {author} {\bibfnamefont {M.}~\bibnamefont
  {Enderlein}}, \bibinfo {author} {\bibfnamefont {T.}~\bibnamefont {Huber}}, \
  and\ \bibinfo {author} {\bibfnamefont {T.}~\bibnamefont {Schaetz}},\ }\href
  {\doibase 10.1038/nphoton.2010.236} {\bibfield  {journal} {\bibinfo
  {journal} {Nat. Photonics}\ }\textbf {\bibinfo {volume} {4}},\ \bibinfo
  {pages} {772} (\bibinfo {year} {2010})}\BibitemShut {NoStop}%
\bibitem [{\citenamefont {Smith}\ \emph {et~al.}(2005)\citenamefont {Smith},
  \citenamefont {Makarov},\ and\ \citenamefont {Lin}}]{Smith2005}%
  \BibitemOpen
  \bibfield  {author} {\bibinfo {author} {\bibfnamefont {W.~W.}\ \bibnamefont
  {Smith}}, \bibinfo {author} {\bibfnamefont {O.~P.}\ \bibnamefont {Makarov}},
  \ and\ \bibinfo {author} {\bibfnamefont {J.}~\bibnamefont {Lin}},\ }\href
  {\doibase 10.1080/09500340500275850} {\bibfield  {journal} {\bibinfo
  {journal} {J. Mod. Opt.}\ }\textbf {\bibinfo {volume} {52}},\ \bibinfo
  {pages} {2253} (\bibinfo {year} {2005})}\BibitemShut {NoStop}%
\bibitem [{\citenamefont {Deiglmayr}\ \emph {et~al.}(2012)\citenamefont
  {Deiglmayr}, \citenamefont {G\"{o}ritz}, \citenamefont {Best}, \citenamefont
  {Weidem\"{u}ller},\ and\ \citenamefont {Wester}}]{Deiglmayr2012}%
  \BibitemOpen
  \bibfield  {author} {\bibinfo {author} {\bibfnamefont {J.}~\bibnamefont
  {Deiglmayr}}, \bibinfo {author} {\bibfnamefont {A.}~\bibnamefont
  {G\"{o}ritz}}, \bibinfo {author} {\bibfnamefont {T.}~\bibnamefont {Best}},
  \bibinfo {author} {\bibfnamefont {M.}~\bibnamefont {Weidem\"{u}ller}}, \ and\
  \bibinfo {author} {\bibfnamefont {R.}~\bibnamefont {Wester}},\ }\href
  {\doibase 10.1103/PhysRevA.86.043438} {\bibfield  {journal} {\bibinfo
  {journal} {Phys. Rev. A}\ }\textbf {\bibinfo {volume} {86}},\ \bibinfo
  {pages} {043438} (\bibinfo {year} {2012})}\BibitemShut {NoStop}%
\bibitem [{\citenamefont {Grier}\ \emph {et~al.}(2009)\citenamefont {Grier},
  \citenamefont {Cetina}, \citenamefont {Oru\v{c}evi\'{c}},\ and\ \citenamefont
  {Vuleti\'{c}}}]{Grier2009}%
  \BibitemOpen
  \bibfield  {author} {\bibinfo {author} {\bibfnamefont {A.~T.}\ \bibnamefont
  {Grier}}, \bibinfo {author} {\bibfnamefont {M.}~\bibnamefont {Cetina}},
  \bibinfo {author} {\bibfnamefont {F.}~\bibnamefont {Oru\v{c}evi\'{c}}}, \
  and\ \bibinfo {author} {\bibfnamefont {V.}~\bibnamefont {Vuleti\'{c}}},\
  }\href {\doibase 10.1103/PhysRevLett.102.223201} {\bibfield  {journal}
  {\bibinfo  {journal} {Phys. Rev. Lett.}\ }\textbf {\bibinfo {volume} {102}},\
  \bibinfo {pages} {223201} (\bibinfo {year} {2009})}\BibitemShut {NoStop}%
\bibitem [{\citenamefont {Smith}\ \emph {et~al.}(2013)\citenamefont {Smith},
  \citenamefont {Goodman}, \citenamefont {Sivarajah}, \citenamefont {Wells},
  \citenamefont {Banerjee}, \citenamefont {C\^{o}t\'{e}}, \citenamefont
  {Michels}, \citenamefont {Mongtomery},\ and\ \citenamefont
  {Narducci}}]{Smith2013}%
  \BibitemOpen
  \bibfield  {author} {\bibinfo {author} {\bibfnamefont {W.~W.}\ \bibnamefont
  {Smith}}, \bibinfo {author} {\bibfnamefont {D.~S.}\ \bibnamefont {Goodman}},
  \bibinfo {author} {\bibfnamefont {I.}~\bibnamefont {Sivarajah}}, \bibinfo
  {author} {\bibfnamefont {J.~E.}\ \bibnamefont {Wells}}, \bibinfo {author}
  {\bibfnamefont {S.}~\bibnamefont {Banerjee}}, \bibinfo {author}
  {\bibfnamefont {R.}~\bibnamefont {C\^{o}t\'{e}}}, \bibinfo {author}
  {\bibfnamefont {H.~H.}\ \bibnamefont {Michels}}, \bibinfo {author}
  {\bibfnamefont {J.~A.}\ \bibnamefont {Mongtomery}}, \ and\ \bibinfo {author}
  {\bibfnamefont {F.~A.}\ \bibnamefont {Narducci}},\ }\href {\doibase
  10.1007/s00340-013-5672-2} {\bibfield  {journal} {\bibinfo  {journal} {Appl.
  Phys. B}\ }\textbf {\bibinfo {volume} {114}},\ \bibinfo {pages} {75}
  (\bibinfo {year} {2013})}\BibitemShut {NoStop}%
\bibitem [{\citenamefont {Schmid}\ \emph {et~al.}(2010)\citenamefont {Schmid},
  \citenamefont {H\"{a}rter},\ and\ \citenamefont {Denschlag}}]{Schmid2010a}%
  \BibitemOpen
  \bibfield  {author} {\bibinfo {author} {\bibfnamefont {S.}~\bibnamefont
  {Schmid}}, \bibinfo {author} {\bibfnamefont {A.}~\bibnamefont {H\"{a}rter}},
  \ and\ \bibinfo {author} {\bibfnamefont {J.~H.}\ \bibnamefont {Denschlag}},\
  }\href {\doibase 10.1103/PhysRevLett.105.133202} {\bibfield  {journal}
  {\bibinfo  {journal} {Phys. Rev. Lett.}\ }\textbf {\bibinfo {volume} {105}},\
  \bibinfo {pages} {133202} (\bibinfo {year} {2010})}\BibitemShut {NoStop}%
\bibitem [{\citenamefont {Zipkes}\ \emph {et~al.}(2011)\citenamefont {Zipkes},
  \citenamefont {Ratschbacher}, \citenamefont {Palzer}, \citenamefont {Sias},\
  and\ \citenamefont {K\"{o}hl}}]{Zipkes2011}%
  \BibitemOpen
  \bibfield  {author} {\bibinfo {author} {\bibfnamefont {C.}~\bibnamefont
  {Zipkes}}, \bibinfo {author} {\bibfnamefont {L.}~\bibnamefont
  {Ratschbacher}}, \bibinfo {author} {\bibfnamefont {S.}~\bibnamefont
  {Palzer}}, \bibinfo {author} {\bibfnamefont {C.}~\bibnamefont {Sias}}, \ and\
  \bibinfo {author} {\bibfnamefont {M.}~\bibnamefont {K\"{o}hl}},\ }\href
  {\doibase 10.1088/1742-6596/264/1/012019} {\bibfield  {journal} {\bibinfo
  {journal} {J. Phys. Conf. Ser.}\ }\textbf {\bibinfo {volume} {264}},\
  \bibinfo {pages} {012019} (\bibinfo {year} {2011})}\BibitemShut {NoStop}%
\bibitem [{\citenamefont {Haze}\ \emph {et~al.}(2013)\citenamefont {Haze},
  \citenamefont {Hata}, \citenamefont {Fujinaga},\ and\ \citenamefont
  {Mukaiyama}}]{Haze2013}%
  \BibitemOpen
  \bibfield  {author} {\bibinfo {author} {\bibfnamefont {S.}~\bibnamefont
  {Haze}}, \bibinfo {author} {\bibfnamefont {S.}~\bibnamefont {Hata}}, \bibinfo
  {author} {\bibfnamefont {M.}~\bibnamefont {Fujinaga}}, \ and\ \bibinfo
  {author} {\bibfnamefont {T.}~\bibnamefont {Mukaiyama}},\ }\href {\doibase
  10.1103/PhysRevA.87.052715} {\bibfield  {journal} {\bibinfo  {journal} {Phys.
  Rev. A}\ }\textbf {\bibinfo {volume} {87}},\ \bibinfo {pages} {052715}
  (\bibinfo {year} {2013})}\BibitemShut {NoStop}%
\bibitem [{\citenamefont {Hall}\ and\ \citenamefont
  {Willitsch}(2012)}]{Hall2012}%
  \BibitemOpen
  \bibfield  {author} {\bibinfo {author} {\bibfnamefont {F.~H.~J.}\
  \bibnamefont {Hall}}\ and\ \bibinfo {author} {\bibfnamefont {S.}~\bibnamefont
  {Willitsch}},\ }\href {\doibase 10.1103/PhysRevLett.109.233202} {\bibfield
  {journal} {\bibinfo  {journal} {Phys. Rev. Lett.}\ }\textbf {\bibinfo
  {volume} {109}},\ \bibinfo {pages} {233202} (\bibinfo {year}
  {2012})}\BibitemShut {NoStop}%
\bibitem [{\citenamefont {Ratschbacher}\ \emph {et~al.}(2012)\citenamefont
  {Ratschbacher}, \citenamefont {Zipkes}, \citenamefont {Sias},\ and\
  \citenamefont {K\"{o}hl}}]{Ratschbacher2012}%
  \BibitemOpen
  \bibfield  {author} {\bibinfo {author} {\bibfnamefont {L.}~\bibnamefont
  {Ratschbacher}}, \bibinfo {author} {\bibfnamefont {C.}~\bibnamefont
  {Zipkes}}, \bibinfo {author} {\bibfnamefont {C.}~\bibnamefont {Sias}}, \ and\
  \bibinfo {author} {\bibfnamefont {M.}~\bibnamefont {K\"{o}hl}},\ }\href
  {\doibase 10.1038/nphys2373} {\bibfield  {journal} {\bibinfo  {journal} {Nat.
  Phys.}\ }\textbf {\bibinfo {volume} {8}},\ \bibinfo {pages} {649} (\bibinfo
  {year} {2012})}\BibitemShut {NoStop}%
\bibitem [{\citenamefont {Hall}\ \emph {et~al.}(2013)\citenamefont {Hall},
  \citenamefont {Eberle}, \citenamefont {Hegi}, \citenamefont {Raoult},
  \citenamefont {Aymar}, \citenamefont {Dulieu},\ and\ \citenamefont
  {Willitsch}}]{Hall2013}%
  \BibitemOpen
  \bibfield  {author} {\bibinfo {author} {\bibfnamefont {F.~H.}\ \bibnamefont
  {Hall}}, \bibinfo {author} {\bibfnamefont {P.}~\bibnamefont {Eberle}},
  \bibinfo {author} {\bibfnamefont {G.}~\bibnamefont {Hegi}}, \bibinfo {author}
  {\bibfnamefont {M.}~\bibnamefont {Raoult}}, \bibinfo {author} {\bibfnamefont
  {M.}~\bibnamefont {Aymar}}, \bibinfo {author} {\bibfnamefont
  {O.}~\bibnamefont {Dulieu}}, \ and\ \bibinfo {author} {\bibfnamefont
  {S.}~\bibnamefont {Willitsch}},\ }\href {\doibase
  10.1080/00268976.2013.780107} {\bibfield  {journal} {\bibinfo  {journal}
  {Mol. Phys.}\ }\textbf {\bibinfo {volume} {111}},\ \bibinfo {pages} {2020}
  (\bibinfo {year} {2013})}\BibitemShut {NoStop}%
\bibitem [{\citenamefont {Willitsch}\ \emph {et~al.}(2008)\citenamefont
  {Willitsch}, \citenamefont {Bell}, \citenamefont {Gingell},\ and\
  \citenamefont {Softley}}]{Willitsch2008}%
  \BibitemOpen
  \bibfield  {author} {\bibinfo {author} {\bibfnamefont {S.}~\bibnamefont
  {Willitsch}}, \bibinfo {author} {\bibfnamefont {M.~T.}\ \bibnamefont {Bell}},
  \bibinfo {author} {\bibfnamefont {A.~D.}\ \bibnamefont {Gingell}}, \ and\
  \bibinfo {author} {\bibfnamefont {T.~P.}\ \bibnamefont {Softley}},\ }\href
  {\doibase 10.1039/b813408c} {\bibfield  {journal} {\bibinfo  {journal} {Phys.
  Chem. Chem. Phys.}\ }\textbf {\bibinfo {volume} {10}},\ \bibinfo {pages}
  {7200} (\bibinfo {year} {2008})}\BibitemShut {NoStop}%
\bibitem [{\citenamefont {Ravi}\ \emph {et~al.}(2012)\citenamefont {Ravi},
  \citenamefont {Lee}, \citenamefont {Sharma}, \citenamefont {Werth},\ and\
  \citenamefont {Rangwala}}]{Ravi2012}%
  \BibitemOpen
  \bibfield  {author} {\bibinfo {author} {\bibfnamefont {K.}~\bibnamefont
  {Ravi}}, \bibinfo {author} {\bibfnamefont {S.}~\bibnamefont {Lee}}, \bibinfo
  {author} {\bibfnamefont {A.}~\bibnamefont {Sharma}}, \bibinfo {author}
  {\bibfnamefont {G.}~\bibnamefont {Werth}}, \ and\ \bibinfo {author}
  {\bibfnamefont {S.~A.}\ \bibnamefont {Rangwala}},\ }\href {\doibase
  10.1038/ncomms2131} {\bibfield  {journal} {\bibinfo  {journal} {Nat.
  Commun.}\ }\textbf {\bibinfo {volume} {3}},\ \bibinfo {pages} {1126}
  (\bibinfo {year} {2012})}\BibitemShut {NoStop}%
\bibitem [{\citenamefont {Doerk}\ \emph {et~al.}(2010)\citenamefont {Doerk},
  \citenamefont {Idziaszek},\ and\ \citenamefont {Calarco}}]{Doerk2010}%
  \BibitemOpen
  \bibfield  {author} {\bibinfo {author} {\bibfnamefont {H.}~\bibnamefont
  {Doerk}}, \bibinfo {author} {\bibfnamefont {Z.}~\bibnamefont {Idziaszek}}, \
  and\ \bibinfo {author} {\bibfnamefont {T.}~\bibnamefont {Calarco}},\ }\href
  {\doibase 10.1103/PhysRevA.81.012708} {\bibfield  {journal} {\bibinfo
  {journal} {Phys. Rev. A}\ }\textbf {\bibinfo {volume} {81}},\ \bibinfo
  {pages} {012708} (\bibinfo {year} {2010})}\BibitemShut {NoStop}%
\bibitem [{\citenamefont {Gerritsma}\ \emph {et~al.}(2012)\citenamefont
  {Gerritsma}, \citenamefont {Negretti}, \citenamefont {Doerk}, \citenamefont
  {Idziaszek}, \citenamefont {Calarco},\ and\ \citenamefont
  {Schmidt-Kaler}}]{Gerritsma2012}%
  \BibitemOpen
  \bibfield  {author} {\bibinfo {author} {\bibfnamefont {R.}~\bibnamefont
  {Gerritsma}}, \bibinfo {author} {\bibfnamefont {A.}~\bibnamefont {Negretti}},
  \bibinfo {author} {\bibfnamefont {H.}~\bibnamefont {Doerk}}, \bibinfo
  {author} {\bibfnamefont {Z.}~\bibnamefont {Idziaszek}}, \bibinfo {author}
  {\bibfnamefont {T.}~\bibnamefont {Calarco}}, \ and\ \bibinfo {author}
  {\bibfnamefont {F.}~\bibnamefont {Schmidt-Kaler}},\ }\href {\doibase
  10.1103/PhysRevLett.109.080402} {\bibfield  {journal} {\bibinfo  {journal}
  {Phys. Rev. Lett.}\ }\textbf {\bibinfo {volume} {109}},\ \bibinfo {pages}
  {080402} (\bibinfo {year} {2012})}\BibitemShut {NoStop}%
\bibitem [{\citenamefont {Joger}\ \emph {et~al.}(2014)\citenamefont {Joger},
  \citenamefont {Negretti},\ and\ \citenamefont {Gerritsma}}]{Joger2014}%
  \BibitemOpen
  \bibfield  {author} {\bibinfo {author} {\bibfnamefont {J.}~\bibnamefont
  {Joger}}, \bibinfo {author} {\bibfnamefont {A.}~\bibnamefont {Negretti}}, \
  and\ \bibinfo {author} {\bibfnamefont {R.}~\bibnamefont {Gerritsma}},\ }\href
  {\doibase 10.1103/PhysRevA.89.063621} {\bibfield  {journal} {\bibinfo
  {journal} {Phys. Rev. A}\ }\textbf {\bibinfo {volume} {89}},\ \bibinfo
  {pages} {063621} (\bibinfo {year} {2014})}\BibitemShut {NoStop}%
\bibitem [{\citenamefont {Ratschbacher}\ \emph {et~al.}(2013)\citenamefont
  {Ratschbacher}, \citenamefont {Sias}, \citenamefont {Carcagni}, \citenamefont
  {Silver}, \citenamefont {Zipkes},\ and\ \citenamefont
  {K\"{o}hl}}]{Ratschbacher2013}%
  \BibitemOpen
  \bibfield  {author} {\bibinfo {author} {\bibfnamefont {L.}~\bibnamefont
  {Ratschbacher}}, \bibinfo {author} {\bibfnamefont {C.}~\bibnamefont {Sias}},
  \bibinfo {author} {\bibfnamefont {L.}~\bibnamefont {Carcagni}}, \bibinfo
  {author} {\bibfnamefont {J.~M.}\ \bibnamefont {Silver}}, \bibinfo {author}
  {\bibfnamefont {C.}~\bibnamefont {Zipkes}}, \ and\ \bibinfo {author}
  {\bibfnamefont {M.}~\bibnamefont {K\"{o}hl}},\ }\href {\doibase
  10.1103/PhysRevLett.110.160402} {\bibfield  {journal} {\bibinfo  {journal}
  {Phys. Rev. Lett.}\ }\textbf {\bibinfo {volume} {110}},\ \bibinfo {pages}
  {160402} (\bibinfo {year} {2013})}\BibitemShut {NoStop}%
\bibitem [{\citenamefont {Casteels}\ \emph {et~al.}(2010)\citenamefont
  {Casteels}, \citenamefont {Tempere},\ and\ \citenamefont
  {Devreese}}]{Casteels2010}%
  \BibitemOpen
  \bibfield  {author} {\bibinfo {author} {\bibfnamefont {W.}~\bibnamefont
  {Casteels}}, \bibinfo {author} {\bibfnamefont {J.}~\bibnamefont {Tempere}}, \
  and\ \bibinfo {author} {\bibfnamefont {J.~T.}\ \bibnamefont {Devreese}},\
  }\href {\doibase 10.1007/s10909-010-0286-0} {\bibfield  {journal} {\bibinfo
  {journal} {J. Low Temp. Phys.}\ }\textbf {\bibinfo {volume} {162}},\ \bibinfo
  {pages} {266} (\bibinfo {year} {2010})}\BibitemShut {NoStop}%
\bibitem [{\citenamefont {Bissbort}\ \emph {et~al.}(2013)\citenamefont
  {Bissbort}, \citenamefont {Cocks}, \citenamefont {Negretti}, \citenamefont
  {Idziaszek}, \citenamefont {Calarco}, \citenamefont {Schmidt-Kaler},
  \citenamefont {Hofstetter},\ and\ \citenamefont {Gerritsma}}]{Bissbort2013}%
  \BibitemOpen
  \bibfield  {author} {\bibinfo {author} {\bibfnamefont {U.}~\bibnamefont
  {Bissbort}}, \bibinfo {author} {\bibfnamefont {D.}~\bibnamefont {Cocks}},
  \bibinfo {author} {\bibfnamefont {A.}~\bibnamefont {Negretti}}, \bibinfo
  {author} {\bibfnamefont {Z.}~\bibnamefont {Idziaszek}}, \bibinfo {author}
  {\bibfnamefont {T.}~\bibnamefont {Calarco}}, \bibinfo {author} {\bibfnamefont
  {F.}~\bibnamefont {Schmidt-Kaler}}, \bibinfo {author} {\bibfnamefont
  {W.}~\bibnamefont {Hofstetter}}, \ and\ \bibinfo {author} {\bibfnamefont
  {R.}~\bibnamefont {Gerritsma}},\ }\href {\doibase
  10.1103/PhysRevLett.111.080501} {\bibfield  {journal} {\bibinfo  {journal}
  {Phys. Rev. Lett.}\ }\textbf {\bibinfo {volume} {111}},\ \bibinfo {pages}
  {080501} (\bibinfo {year} {2013})}\BibitemShut {NoStop}%
\bibitem [{\citenamefont {C\^{o}t\'{e}}(2000)}]{Cote2000a}%
  \BibitemOpen
  \bibfield  {author} {\bibinfo {author} {\bibfnamefont {R.}~\bibnamefont
  {C\^{o}t\'{e}}},\ }\href {\doibase 10.1103/PhysRevLett.85.5316} {\bibfield
  {journal} {\bibinfo  {journal} {Phys. Rev. Lett.}\ }\textbf {\bibinfo
  {volume} {85}},\ \bibinfo {pages} {5316} (\bibinfo {year}
  {2000})}\BibitemShut {NoStop}%
\bibitem [{\citenamefont {H\"{a}rter}\ \emph {et~al.}(2012)\citenamefont
  {H\"{a}rter}, \citenamefont {Kr\"{u}kow}, \citenamefont {Brunner},
  \citenamefont {Schnitzler}, \citenamefont {Schmid},\ and\ \citenamefont
  {Denschlag}}]{Harter2012a}%
  \BibitemOpen
  \bibfield  {author} {\bibinfo {author} {\bibfnamefont {A.}~\bibnamefont
  {H\"{a}rter}}, \bibinfo {author} {\bibfnamefont {A.}~\bibnamefont
  {Kr\"{u}kow}}, \bibinfo {author} {\bibfnamefont {A.}~\bibnamefont {Brunner}},
  \bibinfo {author} {\bibfnamefont {W.}~\bibnamefont {Schnitzler}}, \bibinfo
  {author} {\bibfnamefont {S.}~\bibnamefont {Schmid}}, \ and\ \bibinfo {author}
  {\bibfnamefont {J.~H.}\ \bibnamefont {Denschlag}},\ }\href {\doibase
  10.1103/PhysRevLett.109.123201} {\bibfield  {journal} {\bibinfo  {journal}
  {Phys. Rev. Lett.}\ }\textbf {\bibinfo {volume} {109}},\ \bibinfo {pages}
  {123201} (\bibinfo {year} {2012})}\BibitemShut {NoStop}%
\bibitem [{\citenamefont {Zipkes}\ \emph {et~al.}(2010)\citenamefont {Zipkes},
  \citenamefont {Palzer}, \citenamefont {Sias},\ and\ \citenamefont
  {K\"{o}hl}}]{Zipkes2010}%
  \BibitemOpen
  \bibfield  {author} {\bibinfo {author} {\bibfnamefont {C.}~\bibnamefont
  {Zipkes}}, \bibinfo {author} {\bibfnamefont {S.}~\bibnamefont {Palzer}},
  \bibinfo {author} {\bibfnamefont {C.}~\bibnamefont {Sias}}, \ and\ \bibinfo
  {author} {\bibfnamefont {M.}~\bibnamefont {K\"{o}hl}},\ }\href {\doibase
  10.1038/nature08865} {\bibfield  {journal} {\bibinfo  {journal} {Nature}\
  }\textbf {\bibinfo {volume} {464}},\ \bibinfo {pages} {388} (\bibinfo {year}
  {2010})}\BibitemShut {NoStop}%
\bibitem [{\citenamefont {C\^{o}t\'{e}}\ \emph {et~al.}(2002)\citenamefont
  {C\^{o}t\'{e}}, \citenamefont {Kharchenko},\ and\ \citenamefont
  {Lukin}}]{Cote2002}%
  \BibitemOpen
  \bibfield  {author} {\bibinfo {author} {\bibfnamefont {R.}~\bibnamefont
  {C\^{o}t\'{e}}}, \bibinfo {author} {\bibfnamefont {V.}~\bibnamefont
  {Kharchenko}}, \ and\ \bibinfo {author} {\bibfnamefont {M.~D.}\ \bibnamefont
  {Lukin}},\ }\href {\doibase 10.1103/PhysRevLett.89.093001} {\bibfield
  {journal} {\bibinfo  {journal} {Phys. Rev. Lett.}\ }\textbf {\bibinfo
  {volume} {89}},\ \bibinfo {pages} {093001} (\bibinfo {year}
  {2002})}\BibitemShut {NoStop}%
\bibitem [{\citenamefont {Goold}\ \emph
  {et~al.}(2010{\natexlab{a}})\citenamefont {Goold}, \citenamefont {Doerk},
  \citenamefont {Idziaszek}, \citenamefont {Calarco},\ and\ \citenamefont
  {Busch}}]{Goold2010}%
  \BibitemOpen
  \bibfield  {author} {\bibinfo {author} {\bibfnamefont {J.}~\bibnamefont
  {Goold}}, \bibinfo {author} {\bibfnamefont {H.}~\bibnamefont {Doerk}},
  \bibinfo {author} {\bibfnamefont {Z.}~\bibnamefont {Idziaszek}}, \bibinfo
  {author} {\bibfnamefont {T.}~\bibnamefont {Calarco}}, \ and\ \bibinfo
  {author} {\bibfnamefont {T.}~\bibnamefont {Busch}},\ }\href {\doibase
  10.1103/PhysRevA.81.041601} {\bibfield  {journal} {\bibinfo  {journal} {Phys.
  Rev. A}\ }\textbf {\bibinfo {volume} {81}},\ \bibinfo {pages} {041601}
  (\bibinfo {year} {2010}{\natexlab{a}})}\BibitemShut {NoStop}%
\bibitem [{\citenamefont {Girardeau}(1960)}]{Girardeau1960}%
  \BibitemOpen
  \bibfield  {author} {\bibinfo {author} {\bibfnamefont {M.}~\bibnamefont
  {Girardeau}},\ }\href {\doibase 10.1063/1.1703687} {\bibfield  {journal}
  {\bibinfo  {journal} {J. Math. Phys.}\ }\textbf {\bibinfo {volume} {1}},\
  \bibinfo {pages} {516} (\bibinfo {year} {1960})}\BibitemShut {NoStop}%
\bibitem [{\citenamefont {Chin}\ \emph {et~al.}(2010)\citenamefont {Chin},
  \citenamefont {Grimm}, \citenamefont {Julienne},\ and\ \citenamefont
  {Tiesinga}}]{Chin2010}%
  \BibitemOpen
  \bibfield  {author} {\bibinfo {author} {\bibfnamefont {C.}~\bibnamefont
  {Chin}}, \bibinfo {author} {\bibfnamefont {R.}~\bibnamefont {Grimm}},
  \bibinfo {author} {\bibfnamefont {P.}~\bibnamefont {Julienne}}, \ and\
  \bibinfo {author} {\bibfnamefont {E.}~\bibnamefont {Tiesinga}},\ }\href
  {\doibase 10.1103/RevModPhys.82.1225} {\bibfield  {journal} {\bibinfo
  {journal} {Rev. Mod. Phys.}\ }\textbf {\bibinfo {volume} {82}},\ \bibinfo
  {pages} {1225} (\bibinfo {year} {2010})}\BibitemShut {NoStop}%
\bibitem [{\citenamefont {Olshanii}(1998)}]{Olshanii1998}%
  \BibitemOpen
  \bibfield  {author} {\bibinfo {author} {\bibfnamefont {M.}~\bibnamefont
  {Olshanii}},\ }\href {\doibase 10.1103/PhysRevLett.81.938} {\bibfield
  {journal} {\bibinfo  {journal} {Phys. Rev. Lett.}\ }\textbf {\bibinfo
  {volume} {81}},\ \bibinfo {pages} {938} (\bibinfo {year} {1998})}\BibitemShut
  {NoStop}%
\bibitem [{\citenamefont {Cao}\ \emph {et~al.}(2013)\citenamefont {Cao},
  \citenamefont {Kr\"{o}nke}, \citenamefont {Vendrell},\ and\ \citenamefont
  {Schmelcher}}]{Cao2013}%
  \BibitemOpen
  \bibfield  {author} {\bibinfo {author} {\bibfnamefont {L.}~\bibnamefont
  {Cao}}, \bibinfo {author} {\bibfnamefont {S.}~\bibnamefont {Kr\"{o}nke}},
  \bibinfo {author} {\bibfnamefont {O.}~\bibnamefont {Vendrell}}, \ and\
  \bibinfo {author} {\bibfnamefont {P.}~\bibnamefont {Schmelcher}},\ }\href
  {\doibase 10.1063/1.4821350} {\bibfield  {journal} {\bibinfo  {journal} {J.
  Chem. Phys.}\ }\textbf {\bibinfo {volume} {139}},\ \bibinfo {pages} {134103}
  (\bibinfo {year} {2013})}\BibitemShut {NoStop}%
\bibitem [{\citenamefont {Kr\"{o}nke}\ \emph {et~al.}(2013)\citenamefont
  {Kr\"{o}nke}, \citenamefont {Cao}, \citenamefont {Vendrell},\ and\
  \citenamefont {Schmelcher}}]{Kronke2013}%
  \BibitemOpen
  \bibfield  {author} {\bibinfo {author} {\bibfnamefont {S.}~\bibnamefont
  {Kr\"{o}nke}}, \bibinfo {author} {\bibfnamefont {L.}~\bibnamefont {Cao}},
  \bibinfo {author} {\bibfnamefont {O.}~\bibnamefont {Vendrell}}, \ and\
  \bibinfo {author} {\bibfnamefont {P.}~\bibnamefont {Schmelcher}},\ }\href
  {\doibase 10.1088/1367-2630/15/6/063018} {\bibfield  {journal} {\bibinfo
  {journal} {New J. Phys.}\ }\textbf {\bibinfo {volume} {15}},\ \bibinfo
  {pages} {063018} (\bibinfo {year} {2013})}\BibitemShut {NoStop}%
\bibitem [{\citenamefont {Meyer}\ \emph {et~al.}(1989)\citenamefont {Meyer},
  \citenamefont {Manthe},\ and\ \citenamefont {Cederbaum}}]{Meyer1989}%
  \BibitemOpen
  \bibfield  {author} {\bibinfo {author} {\bibfnamefont {H.-D.}\ \bibnamefont
  {Meyer}}, \bibinfo {author} {\bibfnamefont {U.}~\bibnamefont {Manthe}}, \
  and\ \bibinfo {author} {\bibfnamefont {L.}~\bibnamefont {Cederbaum}},\
  }\href@noop {} {\bibfield  {journal} {\bibinfo  {journal} {Chem. Phys.
  Lett.}\ }\textbf {\bibinfo {volume} {165}},\ \bibinfo {pages} {73} (\bibinfo
  {year} {1989})}\BibitemShut {NoStop}%
\bibitem [{\citenamefont {Beck}\ \emph {et~al.}(2000)\citenamefont {Beck},
  \citenamefont {J\"{a}ckle}, \citenamefont {Worth},\ and\ \citenamefont
  {Meyer}}]{Beck2000}%
  \BibitemOpen
  \bibfield  {author} {\bibinfo {author} {\bibfnamefont {M.~H.}\ \bibnamefont
  {Beck}}, \bibinfo {author} {\bibfnamefont {A.}~\bibnamefont {J\"{a}ckle}},
  \bibinfo {author} {\bibfnamefont {G.~A.}\ \bibnamefont {Worth}}, \ and\
  \bibinfo {author} {\bibfnamefont {H.-D.}\ \bibnamefont {Meyer}},\ }\href@noop
  {} {\bibfield  {journal} {\bibinfo  {journal} {Phys. Rep.}\ }\textbf
  {\bibinfo {volume} {324}},\ \bibinfo {pages} {1} (\bibinfo {year}
  {2000})}\BibitemShut {NoStop}%
\bibitem [{\citenamefont {Alon}\ \emph {et~al.}(2008)\citenamefont {Alon},
  \citenamefont {Streltsov},\ and\ \citenamefont {Cederbaum}}]{Alon2008}%
  \BibitemOpen
  \bibfield  {author} {\bibinfo {author} {\bibfnamefont {O.~E.}\ \bibnamefont
  {Alon}}, \bibinfo {author} {\bibfnamefont {A.~I.}\ \bibnamefont {Streltsov}},
  \ and\ \bibinfo {author} {\bibfnamefont {L.~S.}\ \bibnamefont {Cederbaum}},\
  }\href {\doibase 10.1103/PhysRevA.77.033613} {\bibfield  {journal} {\bibinfo
  {journal} {Phys. Rev. A}\ }\textbf {\bibinfo {volume} {77}},\ \bibinfo
  {pages} {033613} (\bibinfo {year} {2008})}\BibitemShut {NoStop}%
\bibitem [{\citenamefont {Kato}\ and\ \citenamefont {Kono}(2004)}]{Kato2004}%
  \BibitemOpen
  \bibfield  {author} {\bibinfo {author} {\bibfnamefont {T.}~\bibnamefont
  {Kato}}\ and\ \bibinfo {author} {\bibfnamefont {H.}~\bibnamefont {Kono}},\
  }\href {\doibase 10.1016/j.cplett.2004.05.106} {\bibfield  {journal}
  {\bibinfo  {journal} {Chem. Phys. Lett.}\ }\textbf {\bibinfo {volume}
  {392}},\ \bibinfo {pages} {533} (\bibinfo {year} {2004})}\BibitemShut
  {NoStop}%
\bibitem [{\citenamefont {Idziaszek}\ \emph {et~al.}(2009)\citenamefont
  {Idziaszek}, \citenamefont {Calarco}, \citenamefont {Julienne},\ and\
  \citenamefont {Simoni}}]{Idziaszek2009}%
  \BibitemOpen
  \bibfield  {author} {\bibinfo {author} {\bibfnamefont {Z.}~\bibnamefont
  {Idziaszek}}, \bibinfo {author} {\bibfnamefont {T.}~\bibnamefont {Calarco}},
  \bibinfo {author} {\bibfnamefont {P.~S.}\ \bibnamefont {Julienne}}, \ and\
  \bibinfo {author} {\bibfnamefont {A.}~\bibnamefont {Simoni}},\ }\href
  {\doibase 10.1103/PhysRevA.79.010702} {\bibfield  {journal} {\bibinfo
  {journal} {Phys. Rev. A}\ }\textbf {\bibinfo {volume} {79}},\ \bibinfo
  {pages} {010702} (\bibinfo {year} {2009})}\BibitemShut {NoStop}%
\bibitem [{\citenamefont {Gao}(2010)}]{Gao2010}%
  \BibitemOpen
  \bibfield  {author} {\bibinfo {author} {\bibfnamefont {B.}~\bibnamefont
  {Gao}},\ }\href {\doibase 10.1103/PhysRevLett.104.213201} {\bibfield
  {journal} {\bibinfo  {journal} {Phys. Rev. Lett.}\ }\textbf {\bibinfo
  {volume} {104}},\ \bibinfo {pages} {213201} (\bibinfo {year}
  {2010})}\BibitemShut {NoStop}%
\bibitem [{\citenamefont {Idziaszek}\ \emph {et~al.}(2011)\citenamefont
  {Idziaszek}, \citenamefont {Simoni}, \citenamefont {Calarco},\ and\
  \citenamefont {Julienne}}]{Idziaszek2011a}%
  \BibitemOpen
  \bibfield  {author} {\bibinfo {author} {\bibfnamefont {Z.}~\bibnamefont
  {Idziaszek}}, \bibinfo {author} {\bibfnamefont {A.}~\bibnamefont {Simoni}},
  \bibinfo {author} {\bibfnamefont {T.}~\bibnamefont {Calarco}}, \ and\
  \bibinfo {author} {\bibfnamefont {P.~S.}\ \bibnamefont {Julienne}},\ }\href
  {\doibase 10.1088/1367-2630/13/8/083005} {\bibfield  {journal} {\bibinfo
  {journal} {New J. Phys.}\ }\textbf {\bibinfo {volume} {13}},\ \bibinfo
  {pages} {083005} (\bibinfo {year} {2011})}\BibitemShut {NoStop}%
\bibitem [{\citenamefont {Gao}(2013)}]{Gao2013}%
  \BibitemOpen
  \bibfield  {author} {\bibinfo {author} {\bibfnamefont {B.}~\bibnamefont
  {Gao}},\ }\href {\doibase 10.1103/PhysRevA.88.022701} {\bibfield  {journal}
  {\bibinfo  {journal} {Phys. Rev. A}\ }\textbf {\bibinfo {volume} {88}},\
  \bibinfo {pages} {022701} (\bibinfo {year} {2013})}\BibitemShut {NoStop}%
\bibitem [{\citenamefont {Seaton}\ and\ \citenamefont
  {Steenman-Clark}(1977)}]{Seaton1977}%
  \BibitemOpen
  \bibfield  {author} {\bibinfo {author} {\bibfnamefont {M.~J.}\ \bibnamefont
  {Seaton}}\ and\ \bibinfo {author} {\bibfnamefont {L.}~\bibnamefont
  {Steenman-Clark}},\ }\href {\doibase 10.1088/0022-3700/10/13/017} {\bibfield
  {journal} {\bibinfo  {journal} {J. Phys. B At. Mol. Phys.}\ }\textbf
  {\bibinfo {volume} {10}},\ \bibinfo {pages} {2639} (\bibinfo {year}
  {1977})}\BibitemShut {NoStop}%
\bibitem [{\citenamefont {Viehland}\ and\ \citenamefont
  {Mason}(1994)}]{Viehland1994}%
  \BibitemOpen
  \bibfield  {author} {\bibinfo {author} {\bibfnamefont {L.~A.}\ \bibnamefont
  {Viehland}}\ and\ \bibinfo {author} {\bibfnamefont {E.}~\bibnamefont
  {Mason}},\ }\href {\doibase 10.1016/0009-2614(94)01148-6} {\bibfield
  {journal} {\bibinfo  {journal} {Chem. Phys. Lett.}\ }\textbf {\bibinfo
  {volume} {230}},\ \bibinfo {pages} {61} (\bibinfo {year} {1994})}\BibitemShut
  {NoStop}%
\bibitem [{\citenamefont {C\^{o}t\'{e}}\ and\ \citenamefont
  {Dalgarno}(2000)}]{Cote2000}%
  \BibitemOpen
  \bibfield  {author} {\bibinfo {author} {\bibfnamefont {R.}~\bibnamefont
  {C\^{o}t\'{e}}}\ and\ \bibinfo {author} {\bibfnamefont {A.}~\bibnamefont
  {Dalgarno}},\ }\href {\doibase 10.1103/PhysRevA.62.012709} {\bibfield
  {journal} {\bibinfo  {journal} {Phys. Rev. A}\ }\textbf {\bibinfo {volume}
  {62}},\ \bibinfo {pages} {012709} (\bibinfo {year} {2000})}\BibitemShut
  {NoStop}%
\bibitem [{\citenamefont {Moritz}\ \emph {et~al.}(2003)\citenamefont {Moritz},
  \citenamefont {St\"{o}ferle}, \citenamefont {K\"{o}hl},\ and\ \citenamefont
  {Esslinger}}]{Moritz2003}%
  \BibitemOpen
  \bibfield  {author} {\bibinfo {author} {\bibfnamefont {H.}~\bibnamefont
  {Moritz}}, \bibinfo {author} {\bibfnamefont {T.}~\bibnamefont
  {St\"{o}ferle}}, \bibinfo {author} {\bibfnamefont {M.}~\bibnamefont
  {K\"{o}hl}}, \ and\ \bibinfo {author} {\bibfnamefont {T.}~\bibnamefont
  {Esslinger}},\ }\href {\doibase 10.1103/PhysRevLett.91.250402} {\bibfield
  {journal} {\bibinfo  {journal} {Phys. Rev. Lett.}\ }\textbf {\bibinfo
  {volume} {91}},\ \bibinfo {pages} {250402} (\bibinfo {year}
  {2003})}\BibitemShut {NoStop}%
\bibitem [{\citenamefont {Idziaszek}\ \emph {et~al.}(2007)\citenamefont
  {Idziaszek}, \citenamefont {Calarco},\ and\ \citenamefont
  {Zoller}}]{Idziaszek2007}%
  \BibitemOpen
  \bibfield  {author} {\bibinfo {author} {\bibfnamefont {Z.}~\bibnamefont
  {Idziaszek}}, \bibinfo {author} {\bibfnamefont {T.}~\bibnamefont {Calarco}},
  \ and\ \bibinfo {author} {\bibfnamefont {P.}~\bibnamefont {Zoller}},\ }\href
  {\doibase 10.1103/PhysRevA.76.033409} {\bibfield  {journal} {\bibinfo
  {journal} {Phys. Rev. A}\ }\textbf {\bibinfo {volume} {76}},\ \bibinfo
  {pages} {033409} (\bibinfo {year} {2007})}\BibitemShut {NoStop}%
\bibitem [{\citenamefont {Greene}\ \emph {et~al.}(1982)\citenamefont {Greene},
  \citenamefont {Rau},\ and\ \citenamefont {Fano}}]{Greene1982}%
  \BibitemOpen
  \bibfield  {author} {\bibinfo {author} {\bibfnamefont {C.~H.}\ \bibnamefont
  {Greene}}, \bibinfo {author} {\bibfnamefont {A.~R.~P.}\ \bibnamefont {Rau}},
  \ and\ \bibinfo {author} {\bibfnamefont {U.}~\bibnamefont {Fano}},\ }\href
  {\doibase 10.1103/PhysRevA.26.2441} {\bibfield  {journal} {\bibinfo
  {journal} {Phys. Rev. A}\ }\textbf {\bibinfo {volume} {26}},\ \bibinfo
  {pages} {2441} (\bibinfo {year} {1982})}\BibitemShut {NoStop}%
\bibitem [{\citenamefont {Seaton}(1983)}]{Seaton1983}%
  \BibitemOpen
  \bibfield  {author} {\bibinfo {author} {\bibfnamefont {M.~J.}\ \bibnamefont
  {Seaton}},\ }\href {\doibase 10.1088/0034-4885/46/2/002} {\bibfield
  {journal} {\bibinfo  {journal} {Reports Prog. Phys.}\ }\textbf {\bibinfo
  {volume} {46}},\ \bibinfo {pages} {167} (\bibinfo {year} {1983})}\BibitemShut
  {NoStop}%
\bibitem [{\citenamefont {Mies}(1984)}]{Mies1984}%
  \BibitemOpen
  \bibfield  {author} {\bibinfo {author} {\bibfnamefont {F.~H.}\ \bibnamefont
  {Mies}},\ }\href {\doibase 10.1063/1.447000} {\bibfield  {journal} {\bibinfo
  {journal} {J. Chem. Phys.}\ }\textbf {\bibinfo {volume} {80}},\ \bibinfo
  {pages} {2514} (\bibinfo {year} {1984})}\BibitemShut {NoStop}%
\bibitem [{\citenamefont {Vogt}\ and\ \citenamefont
  {Wannier}(1954)}]{Vogt1954}%
  \BibitemOpen
  \bibfield  {author} {\bibinfo {author} {\bibfnamefont {E.}~\bibnamefont
  {Vogt}}\ and\ \bibinfo {author} {\bibfnamefont {G.}~\bibnamefont {Wannier}},\
  }\href {\doibase 10.1103/PhysRev.95.1190} {\bibfield  {journal} {\bibinfo
  {journal} {Phys. Rev.}\ }\textbf {\bibinfo {volume} {95}},\ \bibinfo {pages}
  {1190} (\bibinfo {year} {1954})}\BibitemShut {NoStop}%
\bibitem [{\citenamefont {Massignan}\ \emph {et~al.}(2005)\citenamefont
  {Massignan}, \citenamefont {Pethick},\ and\ \citenamefont
  {Smith}}]{Massignan2005}%
  \BibitemOpen
  \bibfield  {author} {\bibinfo {author} {\bibfnamefont {P.}~\bibnamefont
  {Massignan}}, \bibinfo {author} {\bibfnamefont {C.~J.}\ \bibnamefont
  {Pethick}}, \ and\ \bibinfo {author} {\bibfnamefont {H.}~\bibnamefont
  {Smith}},\ }\href {\doibase 10.1103/PhysRevA.71.023606} {\bibfield  {journal}
  {\bibinfo  {journal} {Phys. Rev. A}\ }\textbf {\bibinfo {volume} {71}},\
  \bibinfo {pages} {023606} (\bibinfo {year} {2005})}\BibitemShut {NoStop}%
\bibitem [{\citenamefont {Czuchaj}\ \emph {et~al.}(1987)\citenamefont
  {Czuchaj}, \citenamefont {Sienkiewicz},\ and\ \citenamefont
  {Miklaszewski}}]{Czuchaj1987}%
  \BibitemOpen
  \bibfield  {author} {\bibinfo {author} {\bibfnamefont {E.}~\bibnamefont
  {Czuchaj}}, \bibinfo {author} {\bibfnamefont {J.}~\bibnamefont
  {Sienkiewicz}}, \ and\ \bibinfo {author} {\bibfnamefont {W.}~\bibnamefont
  {Miklaszewski}},\ }\href {\doibase 10.1016/0301-0104(87)80069-1} {\bibfield
  {journal} {\bibinfo  {journal} {Chem. Phys.}\ }\textbf {\bibinfo {volume}
  {116}},\ \bibinfo {pages} {69} (\bibinfo {year} {1987})}\BibitemShut
  {NoStop}%
\bibitem [{\citenamefont {Johnson}(1977)}]{Johnson1977}%
  \BibitemOpen
  \bibfield  {author} {\bibinfo {author} {\bibfnamefont {B.~R.}\ \bibnamefont
  {Johnson}},\ }\href {\doibase 10.1063/1.435384} {\bibfield  {journal}
  {\bibinfo  {journal} {J. Chem. Phys.}\ }\textbf {\bibinfo {volume} {67}},\
  \bibinfo {pages} {4086} (\bibinfo {year} {1977})}\BibitemShut {NoStop}%
\bibitem [{\citenamefont {Nguy\^{e}n}\ \emph {et~al.}(2012)\citenamefont
  {Nguy\^{e}n}, \citenamefont {Kalev}, \citenamefont {Barrett},\ and\
  \citenamefont {Englert}}]{Nguyen2012}%
  \BibitemOpen
  \bibfield  {author} {\bibinfo {author} {\bibfnamefont {L.~H.}\ \bibnamefont
  {Nguy\^{e}n}}, \bibinfo {author} {\bibfnamefont {A.}~\bibnamefont {Kalev}},
  \bibinfo {author} {\bibfnamefont {M.~D.}\ \bibnamefont {Barrett}}, \ and\
  \bibinfo {author} {\bibfnamefont {B.-G.}\ \bibnamefont {Englert}},\ }\href
  {\doibase 10.1103/PhysRevA.85.052718} {\bibfield  {journal} {\bibinfo
  {journal} {Phys. Rev. A}\ }\textbf {\bibinfo {volume} {85}},\ \bibinfo
  {pages} {052718} (\bibinfo {year} {2012})}\BibitemShut {NoStop}%
\bibitem [{Note1()}]{Note1}%
  \BibitemOpen
  \bibinfo {note} {We make this choice only for numerical convenience, as the
  use of the old units introduced in Sec. II would have the effect to introduce
  an overall factor $\mu /m$. Of course, this transformation does not change
  the underlying physics.}\BibitemShut {Stop}%
\bibitem [{\citenamefont {Dirac}(1930)}]{Dirac1930}%
  \BibitemOpen
  \bibfield  {author} {\bibinfo {author} {\bibfnamefont {P.~A.~M.}\
  \bibnamefont {Dirac}},\ }\href {\doibase 10.1017/S0305004100016108}
  {\bibfield  {journal} {\bibinfo  {journal} {Math. Proc. Cambridge Philos.
  Soc.}\ }\textbf {\bibinfo {volume} {26}},\ \bibinfo {pages} {376} (\bibinfo
  {year} {1930})}\BibitemShut {NoStop}%
\bibitem [{\citenamefont {Frenkel}(1934)}]{Frenkel1934}%
  \BibitemOpen
  \bibfield  {author} {\bibinfo {author} {\bibfnamefont {J.}~\bibnamefont
  {Frenkel}},\ }\href
  {http://books.google.de/books/about/Wave\_Mechanics.html?id=ZGcmAAAAMAAJ\&pgis=1}
  {\emph {\bibinfo {title} {{Wave Mechanics}}}},\ \bibinfo {edition} {1st}\
  ed.\ (\bibinfo  {publisher} {Clarendon Press, Oxford},\ \bibinfo {year}
  {1934})\BibitemShut {NoStop}%
\bibitem [{\citenamefont {Kosloff}\ and\ \citenamefont
  {Tal-Ezer}(1986)}]{Kosloff1986}%
  \BibitemOpen
  \bibfield  {author} {\bibinfo {author} {\bibfnamefont {R.}~\bibnamefont
  {Kosloff}}\ and\ \bibinfo {author} {\bibfnamefont {H.}~\bibnamefont
  {Tal-Ezer}},\ }\href@noop {} {\bibfield  {journal} {\bibinfo  {journal}
  {Chem. Phys. Lett.}\ }\textbf {\bibinfo {volume} {127}},\ \bibinfo {pages}
  {223} (\bibinfo {year} {1986})}\BibitemShut {NoStop}%
\bibitem [{\citenamefont {Penrose}\ and\ \citenamefont
  {Onsager}(1956)}]{Penrose1956}%
  \BibitemOpen
  \bibfield  {author} {\bibinfo {author} {\bibfnamefont {O.}~\bibnamefont
  {Penrose}}\ and\ \bibinfo {author} {\bibfnamefont {L.}~\bibnamefont
  {Onsager}},\ }\href {\doibase 10.1103/PhysRev.104.576} {\bibfield  {journal}
  {\bibinfo  {journal} {Phys. Rev.}\ }\textbf {\bibinfo {volume} {104}},\
  \bibinfo {pages} {576} (\bibinfo {year} {1956})}\BibitemShut {NoStop}%
\bibitem [{\citenamefont {Pethick}\ and\ \citenamefont
  {Smith}(2008)}]{Pethick}%
  \BibitemOpen
  \bibfield  {author} {\bibinfo {author} {\bibfnamefont {C.}~\bibnamefont
  {Pethick}}\ and\ \bibinfo {author} {\bibfnamefont {H.}~\bibnamefont
  {Smith}},\ }\href
  {http://www.cambridge.org/ar/academic/subjects/physics/condensed-matter-physics-nanoscience-and-mesoscopic-physics/boseeinstein-condensation-dilute-gases-2nd-edition}
  {\emph {\bibinfo {title} {{Bose-Einstein Condensation Dilute Gases}}}},\
  \bibinfo {edition} {2nd}\ ed.\ (\bibinfo  {publisher} {Cambridge University
  Press},\ \bibinfo {year} {2008})\BibitemShut {NoStop}%
\bibitem [{\citenamefont {Lieb}\ and\ \citenamefont
  {Liniger}(1963)}]{Lieb1963}%
  \BibitemOpen
  \bibfield  {author} {\bibinfo {author} {\bibfnamefont {E.~H.}\ \bibnamefont
  {Lieb}}\ and\ \bibinfo {author} {\bibfnamefont {W.}~\bibnamefont {Liniger}},\
  }\href {\doibase 10.1103/PhysRev.130.1605} {\bibfield  {journal} {\bibinfo
  {journal} {Phys. Rev.}\ }\textbf {\bibinfo {volume} {130}},\ \bibinfo {pages}
  {1605} (\bibinfo {year} {1963})}\BibitemShut {NoStop}%
\bibitem [{\citenamefont {Bongs}\ \emph {et~al.}(2001)\citenamefont {Bongs},
  \citenamefont {Burger}, \citenamefont {Dettmer}, \citenamefont {Hellweg},
  \citenamefont {Arlt}, \citenamefont {Ertmer},\ and\ \citenamefont
  {Sengstock}}]{Bongs2001}%
  \BibitemOpen
  \bibfield  {author} {\bibinfo {author} {\bibfnamefont {K.}~\bibnamefont
  {Bongs}}, \bibinfo {author} {\bibfnamefont {S.}~\bibnamefont {Burger}},
  \bibinfo {author} {\bibfnamefont {S.}~\bibnamefont {Dettmer}}, \bibinfo
  {author} {\bibfnamefont {D.}~\bibnamefont {Hellweg}}, \bibinfo {author}
  {\bibfnamefont {J.}~\bibnamefont {Arlt}}, \bibinfo {author} {\bibfnamefont
  {W.}~\bibnamefont {Ertmer}}, \ and\ \bibinfo {author} {\bibfnamefont
  {K.}~\bibnamefont {Sengstock}},\ }\href {\doibase 10.1103/PhysRevA.63.031602}
  {\bibfield  {journal} {\bibinfo  {journal} {Phys. Rev. A}\ }\textbf {\bibinfo
  {volume} {63}},\ \bibinfo {pages} {031602} (\bibinfo {year}
  {2001})}\BibitemShut {NoStop}%
\bibitem [{\citenamefont {Alon}\ and\ \citenamefont
  {Cederbaum}(2005)}]{Alon2005}%
  \BibitemOpen
  \bibfield  {author} {\bibinfo {author} {\bibfnamefont {O.~E.}\ \bibnamefont
  {Alon}}\ and\ \bibinfo {author} {\bibfnamefont {L.~S.}\ \bibnamefont
  {Cederbaum}},\ }\href {\doibase 10.1103/PhysRevLett.95.140402} {\bibfield
  {journal} {\bibinfo  {journal} {Phys. Rev. Lett.}\ }\textbf {\bibinfo
  {volume} {95}},\ \bibinfo {pages} {140402} (\bibinfo {year}
  {2005})}\BibitemShut {NoStop}%
\bibitem [{\citenamefont {Ernst}\ \emph {et~al.}(2011)\citenamefont {Ernst},
  \citenamefont {Hallwood}, \citenamefont {Gulliksen}, \citenamefont {Meyer},\
  and\ \citenamefont {Brand}}]{Ernst2011}%
  \BibitemOpen
  \bibfield  {author} {\bibinfo {author} {\bibfnamefont {T.}~\bibnamefont
  {Ernst}}, \bibinfo {author} {\bibfnamefont {D.~W.}\ \bibnamefont {Hallwood}},
  \bibinfo {author} {\bibfnamefont {J.}~\bibnamefont {Gulliksen}}, \bibinfo
  {author} {\bibfnamefont {H.-D.}\ \bibnamefont {Meyer}}, \ and\ \bibinfo
  {author} {\bibfnamefont {J.}~\bibnamefont {Brand}},\ }\href {\doibase
  10.1103/PhysRevA.84.023623} {\bibfield  {journal} {\bibinfo  {journal} {Phys.
  Rev. A}\ }\textbf {\bibinfo {volume} {84}},\ \bibinfo {pages} {023623}
  (\bibinfo {year} {2011})}\BibitemShut {NoStop}%
\bibitem [{\citenamefont {Dunjko}\ \emph {et~al.}(2001)\citenamefont {Dunjko},
  \citenamefont {Lorent},\ and\ \citenamefont {Olshanii}}]{Dunjko2001}%
  \BibitemOpen
  \bibfield  {author} {\bibinfo {author} {\bibfnamefont {V.}~\bibnamefont
  {Dunjko}}, \bibinfo {author} {\bibfnamefont {V.}~\bibnamefont {Lorent}}, \
  and\ \bibinfo {author} {\bibfnamefont {M.}~\bibnamefont {Olshanii}},\ }\href
  {\doibase 10.1103/PhysRevLett.86.5413} {\bibfield  {journal} {\bibinfo
  {journal} {Phys. Rev. Lett.}\ }\textbf {\bibinfo {volume} {86}},\ \bibinfo
  {pages} {5413} (\bibinfo {year} {2001})}\BibitemShut {NoStop}%
\bibitem [{\citenamefont {Girardeau}\ \emph {et~al.}(2001)\citenamefont
  {Girardeau}, \citenamefont {Wright},\ and\ \citenamefont
  {Triscari}}]{Girardeau2001}%
  \BibitemOpen
  \bibfield  {author} {\bibinfo {author} {\bibfnamefont {M.~D.}\ \bibnamefont
  {Girardeau}}, \bibinfo {author} {\bibfnamefont {E.~M.}\ \bibnamefont
  {Wright}}, \ and\ \bibinfo {author} {\bibfnamefont {J.~M.}\ \bibnamefont
  {Triscari}},\ }\href {\doibase 10.1103/PhysRevA.63.033601} {\bibfield
  {journal} {\bibinfo  {journal} {Phys. Rev. A}\ }\textbf {\bibinfo {volume}
  {63}},\ \bibinfo {pages} {033601} (\bibinfo {year} {2001})}\BibitemShut
  {NoStop}%
\bibitem [{\citenamefont {Kolomeisky}\ \emph {et~al.}(2000)\citenamefont
  {Kolomeisky}, \citenamefont {Newman}, \citenamefont {Straley},\ and\
  \citenamefont {Qi}}]{Kolomeisky2000}%
  \BibitemOpen
  \bibfield  {author} {\bibinfo {author} {\bibfnamefont {E.~B.}\ \bibnamefont
  {Kolomeisky}}, \bibinfo {author} {\bibfnamefont {T.~J.}\ \bibnamefont
  {Newman}}, \bibinfo {author} {\bibfnamefont {J.~P.}\ \bibnamefont {Straley}},
  \ and\ \bibinfo {author} {\bibfnamefont {X.}~\bibnamefont {Qi}},\ }\href
  {\doibase 10.1103/PhysRevLett.85.1146} {\bibfield  {journal} {\bibinfo
  {journal} {Phys. Rev. Lett.}\ }\textbf {\bibinfo {volume} {85}},\ \bibinfo
  {pages} {1146} (\bibinfo {year} {2000})}\BibitemShut {NoStop}%
\bibitem [{\citenamefont {Minguzzi}\ \emph {et~al.}(2002)\citenamefont
  {Minguzzi}, \citenamefont {Vignolo},\ and\ \citenamefont
  {Tosi}}]{Minguzzi2002}%
  \BibitemOpen
  \bibfield  {author} {\bibinfo {author} {\bibfnamefont {A.}~\bibnamefont
  {Minguzzi}}, \bibinfo {author} {\bibfnamefont {P.}~\bibnamefont {Vignolo}}, \
  and\ \bibinfo {author} {\bibfnamefont {M.}~\bibnamefont {Tosi}},\ }\href
  {\doibase 10.1016/S0375-9601(02)00042-7} {\bibfield  {journal} {\bibinfo
  {journal} {Phys. Lett. A}\ }\textbf {\bibinfo {volume} {294}},\ \bibinfo
  {pages} {222} (\bibinfo {year} {2002})}\BibitemShut {NoStop}%
\bibitem [{\citenamefont {Z\"{o}llner}\ \emph
  {et~al.}(2006{\natexlab{a}})\citenamefont {Z\"{o}llner}, \citenamefont
  {Meyer},\ and\ \citenamefont {Schmelcher}}]{Zollner2006a}%
  \BibitemOpen
  \bibfield  {author} {\bibinfo {author} {\bibfnamefont {S.}~\bibnamefont
  {Z\"{o}llner}}, \bibinfo {author} {\bibfnamefont {H.-D.}\ \bibnamefont
  {Meyer}}, \ and\ \bibinfo {author} {\bibfnamefont {P.}~\bibnamefont
  {Schmelcher}},\ }\href {\doibase 10.1103/PhysRevA.74.063611} {\bibfield
  {journal} {\bibinfo  {journal} {Phys. Rev. A}\ }\textbf {\bibinfo {volume}
  {74}},\ \bibinfo {pages} {063611} (\bibinfo {year}
  {2006}{\natexlab{a}})}\BibitemShut {NoStop}%
\bibitem [{\citenamefont {Z\"{o}llner}\ \emph
  {et~al.}(2006{\natexlab{b}})\citenamefont {Z\"{o}llner}, \citenamefont
  {Meyer},\ and\ \citenamefont {Schmelcher}}]{Zollner2006}%
  \BibitemOpen
  \bibfield  {author} {\bibinfo {author} {\bibfnamefont {S.}~\bibnamefont
  {Z\"{o}llner}}, \bibinfo {author} {\bibfnamefont {H.-D.}\ \bibnamefont
  {Meyer}}, \ and\ \bibinfo {author} {\bibfnamefont {P.}~\bibnamefont
  {Schmelcher}},\ }\href {\doibase 10.1103/PhysRevA.74.053612} {\bibfield
  {journal} {\bibinfo  {journal} {Phys. Rev. A}\ }\textbf {\bibinfo {volume}
  {74}},\ \bibinfo {pages} {053612} (\bibinfo {year}
  {2006}{\natexlab{b}})}\BibitemShut {NoStop}%
\bibitem [{\citenamefont {Lelas}\ \emph {et~al.}(2009)\citenamefont {Lelas},
  \citenamefont {Juki\'{c}},\ and\ \citenamefont {Buljan}}]{Lelas2009}%
  \BibitemOpen
  \bibfield  {author} {\bibinfo {author} {\bibfnamefont {K.}~\bibnamefont
  {Lelas}}, \bibinfo {author} {\bibfnamefont {D.}~\bibnamefont {Juki\'{c}}}, \
  and\ \bibinfo {author} {\bibfnamefont {H.}~\bibnamefont {Buljan}},\ }\href
  {\doibase 10.1103/PhysRevA.80.053617} {\bibfield  {journal} {\bibinfo
  {journal} {Phys. Rev. A}\ }\textbf {\bibinfo {volume} {80}},\ \bibinfo
  {pages} {053617} (\bibinfo {year} {2009})}\BibitemShut {NoStop}%
\bibitem [{\citenamefont {Busch}\ and\ \citenamefont
  {Huyet}(2003)}]{Busch2003}%
  \BibitemOpen
  \bibfield  {author} {\bibinfo {author} {\bibfnamefont {T.}~\bibnamefont
  {Busch}}\ and\ \bibinfo {author} {\bibfnamefont {G.}~\bibnamefont {Huyet}},\
  }\href {\doibase 10.1088/0953-4075/36/12/313} {\bibfield  {journal} {\bibinfo
   {journal} {J. Phys. B At. Mol. Opt. Phys.}\ }\textbf {\bibinfo {volume}
  {36}},\ \bibinfo {pages} {2553} (\bibinfo {year} {2003})}\BibitemShut
  {NoStop}%
\bibitem [{\citenamefont {Goold}\ and\ \citenamefont
  {Busch}(2008)}]{Goold2008}%
  \BibitemOpen
  \bibfield  {author} {\bibinfo {author} {\bibfnamefont {J.}~\bibnamefont
  {Goold}}\ and\ \bibinfo {author} {\bibfnamefont {T.}~\bibnamefont {Busch}},\
  }\href {\doibase 10.1103/PhysRevA.77.063601} {\bibfield  {journal} {\bibinfo
  {journal} {Phys. Rev. A}\ }\textbf {\bibinfo {volume} {77}},\ \bibinfo
  {pages} {063601} (\bibinfo {year} {2008})}\BibitemShut {NoStop}%
\bibitem [{\citenamefont {Goold}\ \emph
  {et~al.}(2010{\natexlab{b}})\citenamefont {Goold}, \citenamefont {Krych},
  \citenamefont {Idziaszek}, \citenamefont {Fogarty},\ and\ \citenamefont
  {Busch}}]{Goold2010a}%
  \BibitemOpen
  \bibfield  {author} {\bibinfo {author} {\bibfnamefont {J.}~\bibnamefont
  {Goold}}, \bibinfo {author} {\bibfnamefont {M.}~\bibnamefont {Krych}},
  \bibinfo {author} {\bibfnamefont {Z.}~\bibnamefont {Idziaszek}}, \bibinfo
  {author} {\bibfnamefont {T.}~\bibnamefont {Fogarty}}, \ and\ \bibinfo
  {author} {\bibfnamefont {T.}~\bibnamefont {Busch}},\ }\href {\doibase
  10.1088/1367-2630/12/9/093041} {\bibfield  {journal} {\bibinfo  {journal}
  {New J. Phys.}\ }\textbf {\bibinfo {volume} {12}},\ \bibinfo {pages} {093041}
  (\bibinfo {year} {2010}{\natexlab{b}})}\BibitemShut {NoStop}%
\bibitem [{\citenamefont {Gangardt}\ and\ \citenamefont
  {Shlyapnikov}(2003)}]{Gangardt2003}%
  \BibitemOpen
  \bibfield  {author} {\bibinfo {author} {\bibfnamefont {D.~M.}\ \bibnamefont
  {Gangardt}}\ and\ \bibinfo {author} {\bibfnamefont {G.~V.}\ \bibnamefont
  {Shlyapnikov}},\ }\href {\doibase 10.1103/PhysRevLett.90.010401} {\bibfield
  {journal} {\bibinfo  {journal} {Phys. Rev. Lett.}\ }\textbf {\bibinfo
  {volume} {90}},\ \bibinfo {pages} {010401} (\bibinfo {year}
  {2003})}\BibitemShut {NoStop}%
\bibitem [{\citenamefont {Serwane}\ \emph {et~al.}(2011)\citenamefont
  {Serwane}, \citenamefont {Z\"{u}rn}, \citenamefont {Lompe}, \citenamefont
  {Ottenstein}, \citenamefont {Wenz},\ and\ \citenamefont
  {Jochim}}]{Serwane2011}%
  \BibitemOpen
  \bibfield  {author} {\bibinfo {author} {\bibfnamefont {F.}~\bibnamefont
  {Serwane}}, \bibinfo {author} {\bibfnamefont {G.}~\bibnamefont {Z\"{u}rn}},
  \bibinfo {author} {\bibfnamefont {T.}~\bibnamefont {Lompe}}, \bibinfo
  {author} {\bibfnamefont {T.~B.}\ \bibnamefont {Ottenstein}}, \bibinfo
  {author} {\bibfnamefont {A.~N.}\ \bibnamefont {Wenz}}, \ and\ \bibinfo
  {author} {\bibfnamefont {S.}~\bibnamefont {Jochim}},\ }\href {\doibase
  10.1126/science.1201351} {\bibfield  {journal} {\bibinfo  {journal}
  {Science}\ }\textbf {\bibinfo {volume} {332}},\ \bibinfo {pages} {336}
  (\bibinfo {year} {2011})}\BibitemShut {NoStop}%
\end{thebibliography}%

%%% End Bibliography
%%%%%%%%%%%%%%%%%%%%%%%%%%%%%%%%%%%%%%%%%%%%%%%%%%%%%%%%%%%%%%%%%%%%%%%%%%%%%%%%%%%%%%%%

%%% End Document
%%%%%%%%%%%%%%%%%%%%%%%%%%%%%%%%%%%%%%%%%%%%%%%%%%%%%%%%%%%%%%%%%%%%%%%%%%%%%%%%%%%%%%%%
\end{document}